\documentclass[journal=jctcce,manuscript=article]{achemso}
\usepackage{algorithm}
\usepackage{algpseudocode}
\usepackage[version=4]{mhchem} 
\usepackage{amsfonts}
\usepackage{amssymb,amsmath}
\usepackage{enumerate}
\usepackage{bbm}
\usepackage{color}
\usepackage{graphicx}
\usepackage{bm} 
\usepackage{siunitx}
\usepackage{mathtools}
\usepackage{ulem}
\usepackage{subcaption}
\usepackage[compat=1.1.0]{tikz-feynman}
\usepackage{upgreek}
\usepackage{siunitx} 
\usepackage[export]{adjustbox}
\usepackage{hyperref} 
\usepackage[noabbrev]{cleveref} 
\crefformat{equation}{Eq.~(#2#1#3)}
\crefformat{figure}{Fig.~#2#1#3}
\crefformat{table}{Table #2#1#3}

\newcommand{\ii}{\mathbbm{i}}

\newcommand{\Bohr}{\mathrm{a_0}}
\newcommand{\ee}{\mathrm{e}}
\newcommand{\change}[1]{{\color{black}{#1}}} 

\definecolor{zbgreen}{RGB}{73,139,140}
\newcommand{\param}[1]{\ensuremath{#1}}

\usepackage{etoolbox} 
\pretocmd{\cite}{~}{}{}
\pretocmd{\ref}{~}{}{}

\usepackage{xparse, xspace}
\NewDocumentCommand\bdgrnn{g}{%
  \textrm{BDG}%
  \IfNoValueTF{#1}{}{$(#1)$}%
  \textrm{--RNN}%
  \xspace%
}

\newcommand{\ket}[1]{|#1 \rangle}

\newcommand{\braket}[1]{\langle #1 \rangle}

\let\Psi\relax
\let\Phi\relax
\let\Psi\varPsi
\let\Phi\varPhi

\author{Zibo Wu}\altaffiliation{Contributed equally to this work.}
\author{Bohan Zhang}\altaffiliation{Contributed equally to this work.}
\author{Wei-Hai Fang}
\author{Zhendong Li}\email{zhendongli@bnu.edu.cn}
\affiliation{Key Laboratory of Theoretical and Computational Photochemistry, Ministry of Education, College of Chemistry, Beijing Normal University, Beijing 100875, China}

\title[\texttt{achemso} demonstration]
{Hybrid tensor network and neural network quantum states
for quantum chemistry}

\begin{document}

\begin{tocentry}
%
%
%
%
%
\vspace{1.5em}
\includegraphics[height=3.3cm]{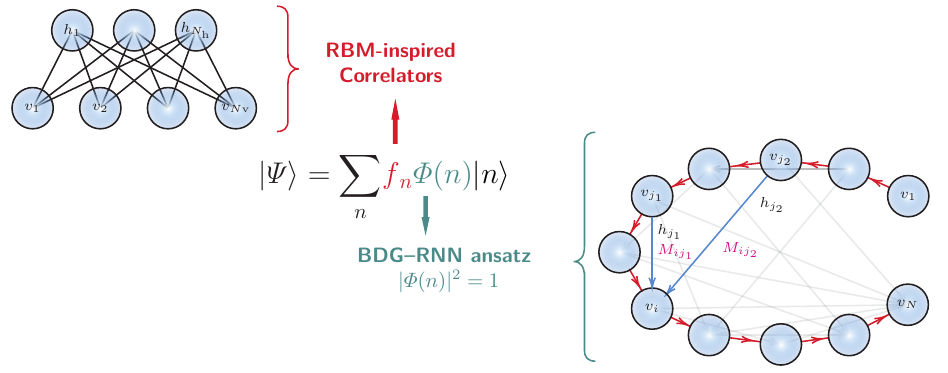}
\end{tocentry}

\begin{abstract}
Neural network quantum states (NQS) have emerged as a powerful and flexible framework for addressing 
quantum many-body problems. While successful for model Hamiltonians, their application to molecular systems remains challenging for several reasons. In this work, we introduce three innovations to overcome some of the key limitations. 
(1) We propose two novel ans\"atze that hybridize tensor network and neural network states for addressing initialization challenges and enhancing the expressivity of tensor networks.
First, we develop a bounded-degree graph recurrent neural network (BDG-RNN) ansatz that leverages graph-based updates, enabling applications to molecular electronic structure problems.
Second, we introduce restricted Boltzmann machine (RBM) inspired correlators
to further enhance expressivity and improve accuracy, without dramatically modifying the underlying variational Monte Carlo (VMC) optimization framework. 
(2) We introduce a semi-stochastic algorithm for local energy evaluation, which significantly reduces computational cost while maintaining high accuracy.
Combining these advances, we demonstrate that our approaches can achieve chemical accuracy in challenging systems, including the one-dimensional hydrogen chain \ce{H50}, the iron–sulfur cluster \ce{[Fe2S2(SCH3)4]^{2-}}, and a three-dimensional $3 \times 3 \times 2$
hydrogen cluster \ce{H18}. These methods are implemented in an open-source package -
\textsc{PyNQS} (https://github.com/Quantum-Chemistry-Group-BNU/PyNQS) to advance 
NQS methodologies for quantum chemistry.
\end{abstract}

\section{Introduction}\label{introduction}
Efficiently solving the electronic Schr\"odinger equation has long been a key challenge for quantum many-body physics and quantum chemistry. 
However, due to the exponential scaling with respect to the electronic degrees of freedom, only a limited number of systems can be solved exactly.
Some standard methods for solving the electronic Schr\"odinger equation include
second-order M{\o}ller–Plesset perturbation (MP2)\cite{Grimme2003_MP2},
configuration interaction (CI)\cite{Werner1988_CI}, coupled cluster (CC)\cite{Bartlett2007_CC}, and density functional theory (DFT)\cite{Kohn1996_DFT}.
In addition to these deterministic methods, quantum Monte Carlo (QMC) approaches provide  effective alternatives through stochastic sampling. 
QMC can be broadly classified into two categories: variational Monte Carlo (VMC)\cite{Ceperley1977_VMC}, which is based on optimizing trial wavefunctions, and
projector Monte Carlo\cite{toulouse2016introduction}, 
including diffusion MC (DMC)\cite{austin2012quantum}, Green's function Monte Carlo (GFMC)\cite{Kalos1962_GFMC, Sorella1998_GFMC}, 
and auxilliary-field QMC (AFQMC)\cite{motta2018ab,lee2022twenty},
which relies on projecting the system's wavefunction onto the ground state through stochastic processes.

Neural network quantum states (NQS) have emerged as a powerful tool to solve quantum many-body problems. This paradigm was first introduced by Carleo and Troyer\cite{Carleo2017}, who demonstrated the use of restricted Boltzmann machines (RBMs) to describe prototypical interacting spin models. NQS leverages the expressive power of deep learning architectures to encode complex entanglement patterns\cite{Hermann2023_reviews}.
Since then, various deep learning architectures, including recurrent neural networks (RNNs)\cite{hibat-allah_recurrent_2020, wu_tensor-network_2023}, convolutional neural networks (CNNs)\cite{Choo2019_CNNS}, 
autoregressive NNs (ARNNs)\cite{Barrett2022,chen2023antn}, Transformers \cite{Zhang2023, Wu2023nnqs, Rende2024}, and retentive networks\cite{knitter2025retentive},
have been integrated into NQS, further extending their applicability to quantum many-body systems. 
Although NQS have demonstrated remarkable progress in quantum many-body physics,
several critical challenges persist in practical applications to molecular
electronic structure problems in quantum chemistry:

(1) One challenge arises from the limitations of NQS ans\"{a}tze.
The great flexibility of NQS without any physical constraint can exhibit a pronounced sensitivity to parameter initialization, particularly in large-scale strongly correlated systems. 
Recent methodological advances have addressed this challenge through diverse pretraining strategies, including the use of configuration interaction 
single and double excitations (CISD) wavefunction\cite{Liu2024} 
and matrix product states (MPS)\cite{Kan2025}.
Another promising direction, which is closely related with
the present study, is to integrate with tensor network states\cite{
wu_tensor-network_2023,chen2023antn,du2025neuralized}
to encoding more physical information into the wavefunction ans\"{a}tze.
For instance, Wu et al.\cite{wu_tensor-network_2023} introduced 
the MPS--RNN ans\"{a}tz, which generalizes RNN to exactly represent MPS
and extended this framework to two-dimensional (2D) systems via a multilinear memory update. However, the complexity of complex molecular systems can significantly exceed the simple 2D representations. 

(2) Efficient optimization strategies remain critical for advancing NQS performance. Recent advances in non-stochastic optimization leverage selected configuration interaction (SCI) methods to deterministically select samples, yielding high-quality molecular energies through RBM\cite{xiang2023, xiang2024}
and neural network backflow (NNBF) ansatz\cite{Liu2024, Liu2025}.

(3) A critical challenge for applying NQS to quantum chemistry
lies in the high computational cost, as local energy evaluations dominate the computational workload in the NQS.
While calculating local energies only within the sample space is a common approximation, this method can become inadequate for strongly correlated systems \cite{Barrett2022, Wu2023nnqs, Zhao2023}. An alternative strategy involves reducing computational complexity through efficient screening of matrix elements\cite{Schwarz2017,Wei2018,Sabzevari2018}.

In this work, we introduce several advancements to overcome some of these limitations. First, building upon the MPS--RNN framework, we propose a bounded--degree graph RNN (\bdgrnn) architecture suitable for molecular electronic structure problems.
Similar to MPS--RNN, BDG--RNN retains the use of MPS initial parameters, which avoids optimization inefficiencies caused by random initialization.
Second, we propose an RBM-inspired correlator architecture to further enhance the accuracy of NQS, while maintaining full compatibility with existing sampling frameworks.
Finally, to reduce high computational costs in large-scale systems, 
we develop an efficient semistochastic scheme for local energy evaluation. The remainder of this article is organized as follows.
In Section \ref{sec:new-ansatz}, we first present an overview of the VMC algorithm, followed by a detailed description of the \bdgrnn ans\"{a}tz, the RBM-inspired correlator, and semistochastic local energy evaluation algorithm.
In Section \ref{sec:detail}, we provide a brief description of the implementation and computational details.
In Section \ref{sec:result}, we apply the proposed methods in prototypical molecular systems, including the 1D hydrogen chain, an active space model of the iron--sulfur cluster \ce{[Fe2S2(SCH3)4]^{2-}}, and the three-dimensional $3\times 3 \times 2$ hydrogen cluster \ce{H18}.
Finally, conclusions and future prospects of the NQS methodologies
are discussed in Section \ref{sec:conlusion}.

\section{Hybrid tensor network and neural network quantum states}
\label{sec:new-ansatz}
\subsection{Overview of VMC and NQS}
Our goal is to solve the electronic Schr\"odinger equation in the second quantization
\begin{equation}
    \hat{H}|\Psi\rangle = E|\Psi\rangle,
\end{equation}
where
\begin{equation}
    \hat{H} = \sum_{pq}h_{pq}\hat{a}_p^{\dagger}\hat{a}_q + \frac{1}{4} \sum_{pqrs}\braket{pq\| rs}\hat{a}_{p}^{\dagger}\hat{a}_q^{\dagger} \hat{a}_s\hat{a}_r,\label{eq:Ham}
\end{equation}
with $h_{pq}$ and $\langle pq\|rs\rangle$ being one-electron
and two-electron molecular integrals, respectively,
$\hat{a}_p^{(\dagger)}$ being the Fermionic
annihilation (creation) operator for the $p$-th
spin orbital. Any quantum state $|\Psi\rangle$ can be expressed as the linear combination of the basis vectors
\begin{equation}
    |\Psi\rangle = \sum_{n_1\cdots n_{2K}} 
    |n_1n_2\cdots n_{2K}\rangle C_{n_1n_2\cdots n_{2K}}
\end{equation}
where $|n_1n_2\cdots n_{2K}\rangle$ represents the occupation number
vectors~(ONV) with $(n_i) \in \{0, 1\}^{2K}$, $K$ is the total number of spatial orbitals, and $C_{n_1n_2\cdots n_{2K}}$ is the corresponding coefficients.
The VMC algorithm is designed to approximate the ground state of a Hamiltonian by optimizing a parametrized trial wavefunction $\Psi_{\theta}$ with parameters $\theta$.
When $\Psi_{\theta}$ is implemented using a neural network, this ansatz is so-called NQS $\ket{\Psi_\theta}$.
The energy $E_\change{\theta}$ of the trial state $\ket{\Psi_\theta}$ can be written as
\begin{equation}
\begin{aligned}
    E_\change{\theta} = 
    \frac{\braket{\Psi_{\theta}|\hat{H}|\Psi_{\theta}}}{\braket{\Psi_{\theta}|\Psi_{\theta}}} 
    = \langle E_{\rm{loc}}(n)\rangle_{n\sim P_\theta(n)},
    \end{aligned}    
\end{equation}
where the probability distribution ${P_{\theta}(n)}$ is defined by
\begin{equation}
    P_{\theta}(n) = \frac{|\braket{\Psi_{\theta}|n}|^2}{\sum_n |\braket{\Psi_{\theta}|n}|^2},
\end{equation}
and $E_{\rm{loc}}(n)$ is the local energy
\begin{equation}
    E_{\rm{loc}}(n) = \sum_m H_{nm}\frac{\Psi_\theta(m)}{\Psi_\theta(n)}.\label{eq:exactElocal}
\end{equation}
The matrix element $H_{nm}$ is defined as $\braket{n|\hat{H}|m}$ where $\ket{n}$ is ONV.
Sampling the ONV $|n\rangle$ according to ${P_{\theta}(n)}$ can be obtained using Markov chain Monte Carlo (MCMC)\cite{Levin2017_Markov} or autoregressive sampling \cite{hibat-allah_recurrent_2020, Barrett2022} for ans\"atze having the autoregressive structure.
The energy gradient with respect to parameters $\theta$ can be evaluated by
\begin{equation}
    \partial_{\theta} E_\change{\theta}  = 2\Re\big[ \big\langle(\partial_\theta\ln{\Psi_\theta^*(n)}) (E_{\rm {loc}}(n) -\langle E\rangle) \big\rangle_{n\sim P_\theta(n)}\big],\label{eq:egrad}
\end{equation}
where $\langle E\rangle$ is a shorthand notation for $\langle E_{\rm{loc}}(n)\rangle_{n\sim P_\theta(n)}$, and $\partial_\theta\ln{\Psi_\theta^*(n)}$ can be calculated using automatic differentiation techniques\cite{Griewank2008_AD}. The parameters $\theta$ are updated according to $\partial_{\theta} E_\change{\theta} $ using an appropriate optimizer, such as stochastic gradient descent~(SGD)\cite{robbins1951stochastic}, Adam\cite{kingma2014adam}, and AdamW\cite{loshchilov2019fixing}.

The accuracy of NQS relies heavily on the expressivity of the variational ansatz. 
The RBM wavefunction \cite{Carleo2017} represents one of the earliest variational NQS ans{\"a}tze 
utilized for characterizing quantum states 
\begin{equation}
    \Psi_\theta(n) = \ee^{\sum_{j=1}^{N_{\rm v}}\param{a}_jn_j} \times \prod_i^{N_h}2\cosh{(\param{b}_i +\sum_{j=1}^{N_{\rm v}}\param{W}_{ij}n_i)},\label{eq:RBM}
\end{equation}
where $\theta = \{(\param{a}_j), (\param{b}_i), (\param{W}_{ij})\}$ are the set of the real or complex network parameters. Here, $N_{\rm h}$ and $N_{\rm v}$ denote the number of hidden and visible units, respectively, and $N_{\rm v}$ typically corresponds to the number of spin orbitals.
The ``hidden variable density" $\alpha$, defined as $ \alpha = N_{\rm h} / N_{\rm v}$, 
can be adjusted to control the accuracy of the wavefunction\cite{Carleo2017, Nomura2017, Choo2020}.

THe RNN wavefunction ansatz has the autoregressive structure\cite{HibatAllah2020}, which
constructs the wavefunction through sequential conditional probability
\begin{equation}
    \Psi(n) = \left(\prod_i\sqrt{P(n_t| n_{<t}^\circ)}\right)\ee^{\ii \phi(n)}
\end{equation}
where $P(n_t| n_{<t}^\circ)$ represents the conditional probability.
For the later convenience, we use $^\circ$ to indicate that the values of $n_{<t}$
are specified during the forward or sampling process.
And $\phi(n)$ represents the phase of the wavefunction.
The architecture operates through two key components at each step $t$:
(1) The memory cell \change{$h_t^{n_t}=\mathrm{RNN}(h_{t-1}^{n_{t-1}^\circ}, n_{t-1}^\circ)$, where
$h_t^{n_t^\circ} \in \mathbb{R}^\chi$ or $\mathbb{C}^\chi$ for fixed $t$ and specified $n_t^\circ$},
updated via RNN units that can encode historical information \change{$h^{n^\circ_{t-1}}_{t-1}$},  
and (2) conditional wavefunction components consisting of amplitude $\sqrt{P(n_t|n^\circ_{<t})}$ and phase $\phi_t$, generated through two functions $f(h_t)$ and $g(h_t)$. For instance, the probability distribution can be computed as $P(n_t|n^\circ_{<t}) = n_t \cdot \operatorname{softmax}(\param{U^{(1)}}h^{\change{n_t}}_t + \param{c^{(1)}})$
and the phase is calculated  as $\phi_t = \uppi \cdot \operatorname{softsign}(\param{U^{(2)}}h^{\change{n_t}}_t + \param{c^{(2)}})$. Finally, the total phase is computed via $\phi(n) = \sum_t \phi_t$. The expressivity of RNN wavefunctions can be modulated by the choice of RNN architectures, such as GRU (Gated Recurrent Unit)\cite{Chung2014} and LSTM (Long Short-Term Memory)\cite{Hochreiter1997}, as well as the form of the functions $f(h_t^{\change{n_t}})$ and $g(h_t^{\change{n_t}})$. By leveraging conditional probabilities, RNNs wavefunction can achieve accurate sampling through autoregressive sampling and avoid the inefficiency of MCMC.

\begin{figure}[!ht]
    \centering
     \includegraphics[width=0.9\textwidth]{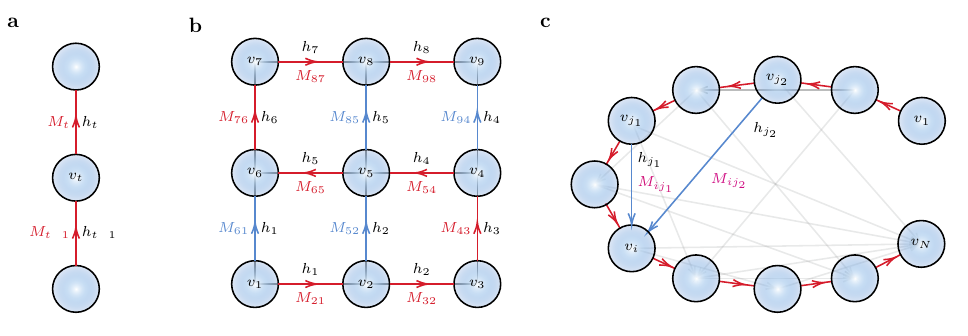}
    \caption{
    Different MPS-RNN wavefunction architectures: (a) the 1D MPS--RNN; (b) the 2D MPS-RNN with horizontal (red arrows) and vertical (blue arrows) memory cell interactions; (c) the \bdgrnn{k} featuring additional non-adjacent memory cell connections (blue arrows), where $k$ denotes the out-degree constraint in bounded-degree graph. Each node corresponds to a spatial orbital,
    and the ordering of spatial orbitals is determined by the Fiedler ordering\cite{barcza2011quantum,olivares2015ab} used in DMRG.   
    Red arrows indicate the ordering of computations.
    }
    \label{fig:rnn-variant}
\end{figure}

Recently, Wu et al.\cite{wu_tensor-network_2023} reformulated the memory cell $h_t^{\change{n_t}}$ update to map MPS to RNN, see Fig. \ref{fig:rnn-variant}a. 
Suppose the tensor of MPS at the site $t$ is denoted by $\param{M_t^{n_t}}$
($(n_t) \in \{0, 1, 2, 3\}^K$), which is a $\chi$-by-$\chi$ real/complex matrix 
\change{for fixed $t$ and specified $n^\circ_t$}. Then, the memory cell $h_t^{n_t}$ is
\begin{equation}
    h_t^{n_t} = \param{M_{t-1}^{n_t}}h^{\change{n_{t-1}^\circ}}_{t-1} + \param{v_t^{n_t}},
    \label{eq:MPS-RNN-1D}
\end{equation}
where $h_t^{n_t}$ and $\param{v_t}^{n_t}$ $\in \mathbb{R}^{\chi} \text{ or } \mathbb{C}^{\chi}$ \change{for fixed $t$ and specified $n_t^\circ$ are vectors.}
The amplitude of wavefunction is defined through\cite{wu_tensor-network_2023}
$P_t^{n_t}= (h_t^{n_t})^\dagger\param{\eta_t} h_t^{n_t}$,
where $\param{\eta_t}$ is a $\chi$-by-$\chi$ non-negative diagonal matrix for fixed $t$.
The phase component $\phi_t$ is computed via an additional linear transformation $\phi_t= \arg{([\param{w_t}]^{\mathrm{T}} h_t^{n^\circ_t}+\param{c_t})}$, where $\param{w_t} \in\mathbb{C}^{\chi}$
and $\param{c_t}\in \mathbb{C}$ for fixed $t$, and
the total phase is computed via $\phi(n)=\sum_t \phi_t$.
For the 2D case, the memory cell update is extended to incorporate interactions along both horizontal and vertical directions, see Fig.\ref{fig:rnn-variant}b.
The memory cell $h_t$ at site $t$ is computed as a combination of contributions from its horizontal $h_{t\text{h}}$ and vertical $h_{t\text{v}}$ neighbouring cells
\begin{equation}
\label{eq:cal_h_mpsrnn2d}
        h_t^{n_t} = \param{M_{t\rm{h}}}^{n_t} h_{t\rm{h}}^{\change{n^\circ_{t\mathrm{h}}}}+ \param{M_{t\rm{v}}^{n_t}} h_{t\rm{v}}^{\change{n^\circ_{t\mathrm{v}}}} + \param{v}_t^{n_t},
\end{equation}
where $\param{M_{t\rm{h}}}^{n_t}$ and $\param{M_{t\rm{v}}}^{n_t}$ are $\chi$-by-$\chi$ parameter matrices acting on the horizontal and vertical directions for fixed $t$ and specified $n_t$, respectively.

\subsection{Bounded--degree graph RNN (BDG--RNN)}
To describe the more complex entanglement among orbitals in molecules,
we generalize MPS-RNN to \bdgrnn, see Fig. \ref{fig:rnn-variant}c, 
where memory cell updates are generalized beyond fixed horizontal/vertical 
dependencies to adaptively incorporate interactions defined by a bounded--degree graph (BDG). To generate the BDG, we first determine the ordering of
orbitals based on the exchange integral $K_{ij}=[ij|ji]$ 
between spatial orbitals $i$ and $j$, using the Fiedler algorithm \cite{barcza2011quantum,olivares2015ab}.
to construct a directed acyclic graph with a maximum out-degree constraint $k = 1$ excluding sources and sinks. Then, we add edges between a given vertex
and other vertices via a greedy algorithm until the maximum out-dedgress $k$
is reached. We summarize the processes in Algorithms S1 and S2, which transform the initial graph $G$ into a modified graph $G^{\prime}$ with an maximum out-degree $k$. Based on the obtained graph $G^{\prime}$, the \bdgrnn parameters can be structured through three fundamental components:
\begin{itemize}
    \item Every vertex $v_i$ maintains a memory cell $h^{n_i}_i$ with assigned parameters $\param{v_i},\param{\eta_i},\param{w_i},\param{c_i}$.
    
    \item Every edge directed $v_j\to v_i$ is assigned a parameter matrix $\param{M_{ij}}^{n_i}$.
    
    \item If vertex $v_i$ receives connections from multiple neighbors $\{v_{\alpha}, v_{\beta}, 
    \dots, v_{\gamma}\}$, a tensor term $\param{T_{i\alpha\beta\dots\gamma}}^{n_i}$ is introduced.
\end{itemize}
Then, the memory cell \change{$h_i^{n_i}$} is defined followed as:
\begin{equation}
\begin{aligned}    
\change{h_i^{n_i}} & =  \operatorname{RNN}\left(\{h^{\change{n^\circ_\alpha}}_\alpha\}_{\alpha \in \mathcal{N}(i)}\right) \\
   & = \sum_{\alpha \in \mathcal{N}(i)} \param{M_{i\alpha}}^{n_i} h_{\alpha}^{\change{n^\circ_\alpha}} + 
   \param{T_{i\alpha\beta\cdots\gamma}}^{n_i}
    \underbrace{h^{\change{n^\circ_\alpha}}_{\alpha}
    h_{\beta}^{\change{n^\circ_\beta}}
    \cdots 
    h_{\gamma}^{\change{n^\circ_\gamma}}}_{d \text{ items}} + \param{v_i}^{n_i}, 
\end{aligned}
\label{eq:bdg-rnn-update}
\end{equation}
where $\mathcal{N}(i)$ denotes the neighborhood of vertex $i$.
The tensor term $\param{T^{n_i}_{i\alpha\beta\cdots\gamma}}$ in Eq . \ref{eq:bdg-rnn-update} requires $\mathcal{O}(\chi^{d})$ parameters, 
where $d$ represents the in-degree of the vertex.
To reduce the computational scaling, we adopt the Tucker decomposition\cite{tucker1966some} employed in Ref. \cite{wu_tensor-network_2023}, viz.,
\begin{equation}
    \param{T^{n_i}_{i\alpha\beta\cdots\gamma}} = \param{U^{n_i}_{i'\alpha'\beta'\cdots\gamma'} 
    \underbrace{K_{ii'}K_{\alpha\alpha'}K_{\beta\beta'}\dots K_{\gamma\gamma'}}}_{d \text{ items}},
\end{equation}
where $\param{U}$ is rank-$(d+1)$ core tensor of shape $(\chi^{\prime}, \dots,\chi^{\prime})$, 
and $\param{K_{ii'}}$ are the corresponding factor matrices of shape $(\chi, \chi^{\prime})$ for each mode. 
We set $\chi^{\prime} = \lceil \chi^{2/(d+1)} \rceil$ to perform the compression, which reduces the parameter complexity from $\mathcal{O}(\chi^d)$ to $\mathcal{O}(d\chi\chi^{\prime} + {\chi^{\prime}}^d)$.
The amplitude $\sqrt{P_t}$ and phase $\phi_t$ components maintain the same computational structures to those in the vanilla MPS--RNN. 
We define two architectural variants based on Eq. \ref{eq:bdg-rnn-update}: 
the \bdgrnn retaining without the tensor terms, and the BDG--TensorRNN incorporating both matrix and tensor terms.
This unified framework allows for systematic transitions between different computational hierarchy:
from 1D MPS--RNN to 2D MPS--RNN and finally to \bdgrnn{k} through controlled modifications of the BDG topology.

\subsection{RBM-inspired correlators}
To further improve the accuracy of NQS, we introduce an correlator in the following way
\begin{equation}
    |\Psi_\theta\rangle = \sum_nf_n\Phi(n)|n\rangle,\quad \sum_n |\Phi(n)|^2=1,
\end{equation}
where $\Phi(n)$ is a normalized NQS, $f_n$ is an RBM-inspired correlator (vide post), and $|\Psi_\theta\rangle$ is not normalized. In this work, we employ two correlators inspired by the form of RBMs \eqref{eq:RBM}, which support
volume-law entanglement\cite{deng2017quantum}.
First, we replace cosh with cos, and omit the exponential term $\ee^{\sum_{j=1}^{N_{\rm v}}\param{a}_jn_j}$, thereby obtaining the following correlator
\begin{equation}
    f_{\text{cos-RBM}}(n) = \prod_{k=1}^{N_{\rm h}}\cos(\param{b}_k + \sum_{i=1}^{N_{\rm v}}\param{W}_{ki}n_i),
\end{equation}
which will be referred to as the cos-RBM correlator.
Second, inspired by the work by Yang et al.\cite{Yanai2020_RBM}, we further introduce
a quadratic term in the correlator,
\begin{equation}
    f_{\text{Ising-RBM}}(n) = \prod_{k=1}^{N_{\rm h}} \cos(\param{b}_k + \sum_{i=1}^{N_{\rm v}}\param{W}_{ki}^{(1)}n_i + \frac{1}{2}\sum_{i,j=1}^{N_{\rm v}}\param{W}_{kij}^{(2)}n_in_j),
\end{equation}
which will be referred to as the Ising-RBM in the later context.
The parameter complexities of the cos-RBM and Ising-RBM scale as 
$\mathcal{O}(\alpha K^2)$ and $\mathcal{O}(\alpha K^3)$, respectively.

The energy for $|\Psi_\theta\rangle$ can be derived 
by reweighting\cite{becca2017quantum} as 
\begin{equation}
    E_\theta = \dfrac{\langle\Psi_\theta|\hat H|\Psi_\theta\rangle}{\langle\Psi_\theta|\Psi_\theta\rangle}
    =\dfrac{\sum_n|\Phi(n)|^2\dfrac{\langle \Psi_\theta | n\rangle \langle n|\hat H|\Psi_\theta\rangle}{|\Phi(n)|^2}}{\sum_n |\Phi(n)|^2|f_n|^2} 
    = \dfrac{\bigg\langle f_n^*\dfrac{\langle n|\hat H|\Psi_\theta\rangle}{\Phi(n)} \bigg\rangle_{n\sim |\Phi(n)|^2}}{\big\langle |f_n|^2\big\rangle_{n\sim |\Phi(n)|^2}}.
\end{equation}
Let $B = \langle |f_n|^2 \rangle_{n\sim |\Phi(n)|^2}$, we can redefine the local energy as
\begin{align}
    \widetilde{E}_{\rm loc}(n) &= \dfrac{f_n^*}{B}\dfrac{\langle n|\hat H|\Psi\rangle}{\Phi(n)}
    = \dfrac{f_n^*}{B} \dfrac{\sum_m\langle n |\hat H| m\rangle \langle m|\Psi\rangle}{\Phi(n)}
    =\sum_m \dfrac{f_n^*}{\sqrt{B}}\dfrac{f_m}{\sqrt{B}}\dfrac{H_{nm}\Phi(m)}{\Phi(n)}\nonumber\\
    &=\sum_m \widetilde{f}_n^*H_{nm}\widetilde{f}_m\dfrac{\Phi(m)}{\Phi(n)},
\end{align}
with $\widetilde{f}_n = f_n/\sqrt{B}$, such that $E_\theta=\langle \widetilde{E}_{\rm loc}(n)\rangle_{n\sim |\Phi(n)|^2}$. The energy gradient can be derived as (see Supporting Information for details)
\begin{equation}
        \partial_\theta E_\theta = 2\Re\big[ \big\langle(\partial_\theta\ln(f_n\Phi(n))^*) (\widetilde{E}_{\rm loc}-|\widetilde{f}_n|^2\langle E\rangle) \big\rangle_{n\sim|\phi(n)|^2} \big].
\end{equation}
Note that when $f_n \equiv 1$, both the energy and its gradient revert to the conventional VMC formulation, see Eq. \eqref{eq:egrad}.
Moreover, the computational cost for evaluating these quantities remains comparable to the original method, with no significant overhead introduced.

\subsection{Semistochastic algorithm for local energy}
The standard algorithm for the local energy $E_{\rm{loc}}(n)$ is via Eq. \eqref{eq:exactElocal},
which for the ab initio Hamiltonian \eqref{eq:Ham} scales as $O(N_{\rm{s}}K^4)$ for computing
nonzero $H_{nm}$ plus $O(N_{\rm{s}}K^4)$ times the computational cost for computing $\Psi(n)$,
with $N_{\rm{s}}$ being the (unique) sample size.
For complex models such as NQS, the second part is dominant. 
Previous works\cite{Barrett2022, Wu2023nnqs, Zhao2023} often use an approximation that only computes the sum over $m$ in the local energy Eq. \eqref{eq:exactElocal} within the sample generated in VMC.
However, this approximation introduces bias, which disappears only when the sample
size goes to infinity.
In this work, we propose a semistochastic algorithm to mainly reduce the computation of the second part. We use the Slater--Condon rules for evaluating
the matrix elements $H_{nm}$, which are less time-consuming than the computation
of the wavefunction, and hence we can decompose the local energy into two components
based on the magnitude of 
$H_{nm}$: the {\it deterministic} part and the {\it stochastic} part. Specifically, the deterministic part $E_{\rm{loc}}^{\rm{d}}(n, \epsilon)$ involves summing over all the matrix elements $H_{nm}$ that satisfy $|H_{nm}| \geq \epsilon$, ensuring that the contributions from these large terms are fully accounted for, viz.,
\begin{equation}
    E_{\rm{loc}}^{\rm{d}}(n, \epsilon) = \sum_{\{ m \,:\, |H_{nm}| \geq \epsilon\}} H_{nm}\frac{\Psi_\theta(m)}{\Psi_\theta(n)}. 
\end{equation}
This part is similar in spirit to that in Ref. \cite{Sabzevari2018} based on screening.
However, only using this part will introduce a bias in the local energy. 

The stochastic part $E_{\rm{loc}}^{\rm{s}}(n, \epsilon, N_{\epsilon})$ is designed to handle the smaller contributions by sampling $m^{\prime}$ from the distribution $ P_n(m^{\prime}) \propto |H_{nm^{\prime}}|$, where $m'$ is defined by $|H_{nm^{\prime}}| < \epsilon$. 
This part helps to reduce the computational cost for smaller terms, which would otherwise be computationally expensive. The number of samples denoted as $N_{\epsilon}$ is chosen based on the desired accuracy. Specifically, we can express the evaluation of the stochastic part by
\begin{equation}
    E_{\rm {loc}}^{\rm{s}}(n, \epsilon, N_{\epsilon}) = 
    \Big\langle 
    \frac{H_{nm^{\prime}}}{P_n(m^{\prime})}\frac{\Psi_\theta(m^{\prime})}{\Psi_\theta{(n)}}
    \Big\rangle_{m^\prime \sim P_n(m')}.
\end{equation}
Thus, the final local energy $E_{\rm{loc}}(n)$ is evaluated as
\begin{equation}
    E_{\rm{loc}}(n, \epsilon, N_{\epsilon}) = E_{\rm{loc}}^{\rm{d}}(n, \epsilon) + E_{\rm loc}^{\rm{s}}(n, \epsilon, N_{\epsilon}), 
\end{equation}
where both $\epsilon$ and $N_{\epsilon}$ are the hyperparameters in this scheme. Unlike
$E_{\rm{loc}}^{\rm{d}}(n, \epsilon)$, $E_{\rm{loc}}(n, \epsilon, N_{\epsilon})$
is an unbiased estimator for the energy. Note that when $\epsilon \to 0 $ or $N_{\epsilon} \to \infty$, $E_{\rm{loc}}(n, \epsilon, N_{\epsilon})$ becomes the exact local energy \eqref{eq:exactElocal}. In practice, it is necessary to choose reasonably $\epsilon$ and $N_{\epsilon}$ in order to strike a balance between the variance and the computational complexity. We will illustrate this point for each molecules examined
in this work.

\section{Implementation and computational details}\label{sec:detail}
We implemented the above ans\"atze and semistochastic algorithm for local energy 
in an open-source package \textsc{PyNQS}\cite{pynqs_github} based on 
the PyTorch deep learning framework\cite{paszke2019pytorch}.
In addition to BDG-RNN, it also supports other commonly used NQS ans\"atze
such as RBM and Transformers, and new ans\"atze can be easily added.
The Hamiltonian matrix elements are evaluated using the Slater--Condon rules and hardware-efficient bitwise operation on graphics processing units (GPUs).
To avoid out of memory, autoregressive sampling if \bdgrnn is implemented via a hybrid depth-first search (DFS) and breadth-first search (BFS) approach, as detailed in Algorithm S3.
The \bdgrnn is initialized by using MPS generated from the \textsc{Focus}\cite{Li2021a, xiang2024distributed} package. The molecular integrals are obtained using the PySCF package\cite{sun2020recent}. All variational parameters are optimized using the AdamW\cite{loshchilov2019fixing} algorithm with first- and second-moments set to $\beta_1=0.9 $ and $\beta_2=0.999$, respectively. In this work, we use double precision (float64) for 
all the numerical data, but the program also supports single precision (float32) for reduced memory usage or improved performance when appropriate.\\

The \bdgrnn supports the use of two types of symmetries, particle number and spin projection, using the autoregressive structures. During the autoregressive sampling process, the $k$-th spin orbitals must satisfy the following constraints:
\begin{equation}
    \begin{aligned}
    N_{\alpha} - \left(\frac{N_e}{2} - \lfloor \frac{k}{2} \rfloor \right) & \leq 
    \sum_{j=0}^{\lfloor \frac{k}{2} \rfloor}x_{2j} \leq N_{\alpha} \\
    N_{\beta} - \left(\frac{N_e}{2} - \lfloor \frac{k}{2} \rfloor \right) & \leq 
    \sum_{j=0}^{\lfloor \frac{k}{2} \rfloor}x_{2j+1} \leq N_{\beta} \\
    \end{aligned}
\end{equation}
where $|x_0x_1\cdots x_{2K-1}x_{2K}\rangle=\ket{n_{1\alpha}n_{1\beta}\dots n_{K\alpha}n_{K\beta}}$, $N_{\alpha}$ and $N_{\beta}$ are the numbers of the $\alpha$ and $\beta$ electrons, respectively.
These physical constraints can be incorporated into the conditional probability distributions as follows\cite{Zhao2023}
\begin{equation}
    \tilde{P}(x_i|x_{i-1}, \dots, x_1) = 
    \begin{cases}
     1, \ \text{if}\ N_{\alpha} - \left(\frac{N_e}{2} - \lfloor \frac{k}{2} \rfloor\right) >
     \sum_{j=0}^{\lfloor \frac{k}{2} \rfloor}x_{2j} & \\
     1, \ \text{if}\ N_{\beta} - \left(\frac{N_e}{2} -\lfloor \frac{k}{2} \rfloor\right) > 
     \sum_{j=0}^{\lfloor \frac{k}{2} \rfloor}x_{2j+1} & \\
     0, \ \text{if}\ N_{\alpha} \leq \sum_{j=0}^{\lfloor \frac{k}{2} \rfloor}x_{2j}  & \\
     0, \ \text{if}\ N_{\beta} \leq \sum_{j=0}^{\lfloor \frac{k}{2} \rfloor}x_{2j+1} & \\
     P(x_i|x_{i-1}, \dots, x_1), \ \text{otherwise} &
    \end{cases}
\end{equation}
Note that after replacing $P(x_i|x_{i-1}, \dots, x_1) $ with $\tilde{P}(x_i|x_{i-1}, \dots, x_1)$, the normalization condition is still satisfied.

\section{Results and discussion}\label{sec:result}
\subsection{Hydrogen chains}
We choose the 1D hydrogen chain as the first benchmark system
to investigate the accuracy of the \bdgrnn and RBM-inspired correlator
as well as the performance of the semistochastic algorithm
for local energy. For this system, MPS offers a compact description 
of ground state entanglement\cite{Schollwoeck2011},
and exact results are available\cite{Hachmann2006}
for comparison. All hydrogen atoms are equally spaced with a bond length of $R =2.0 \Bohr$.
The STO-6G basis set and orthonormalized atomic orbitals (OAOs) are used following the
previous work\cite{Hachmann2006}. 

\begin{figure}[!htp]
    \centering
    \includegraphics[width=0.95\linewidth]{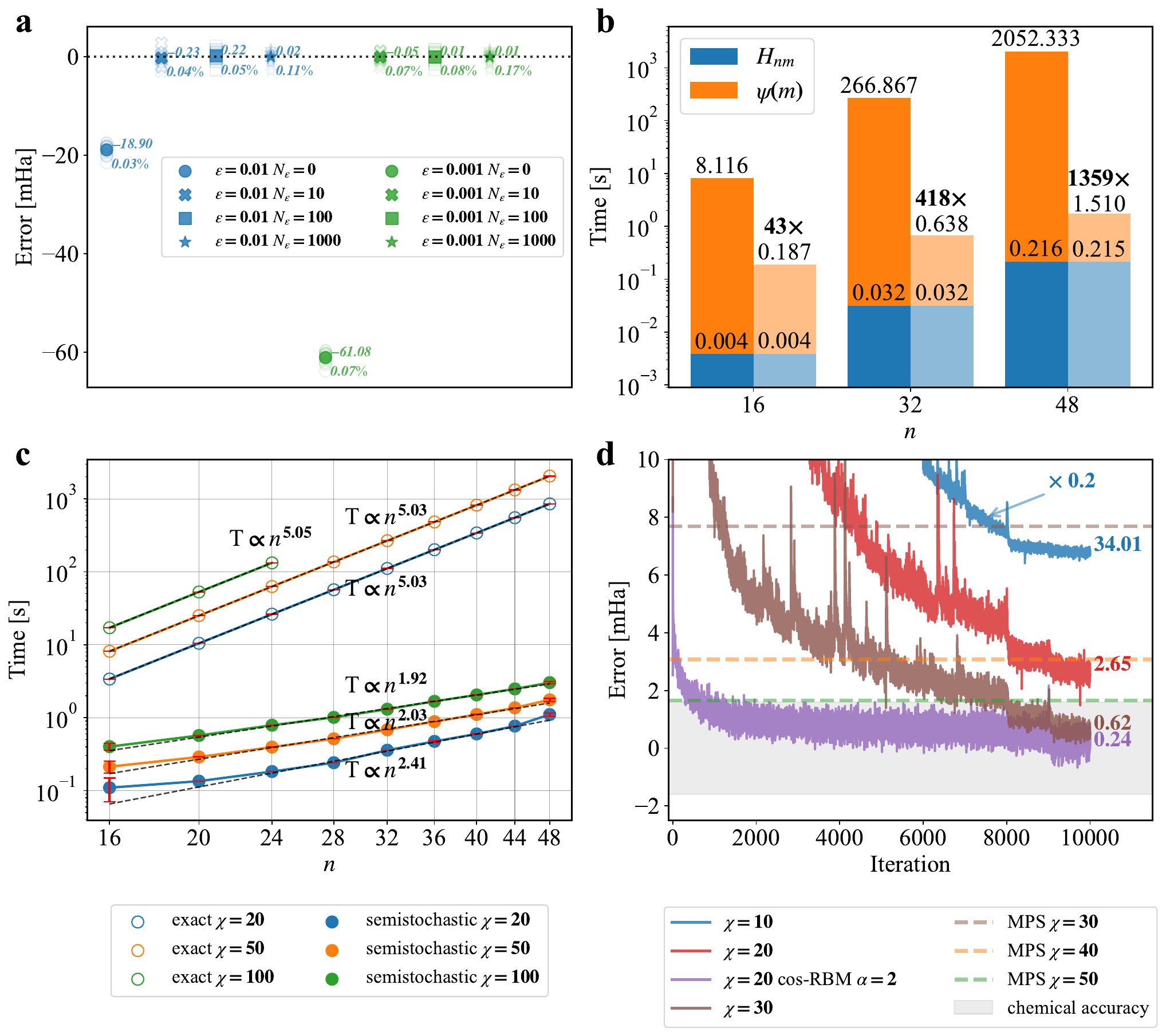}
    \caption{
    Results for the 1D hydrogen chains with a bond length of $2.00 \Bohr$.
    \textbf{a}: Comparison of the performance of the semistochastic algorithm for local energy using different $\epsilon$ and $N_{\epsilon}$ for \ce{H50} using the \bdgrnn{1} ansatz with $\chi = 10$. The percentage of computational cost relative to the original algorithm is shown below each marker, and the error relative to the exact local energy is shown above the marker. 
    Results are obtained using a fixed sample size of $N_s = 5\times 10^4$.
    \textbf{b}:
    Detailed timing for local energy evaluations, including the evaluation of matrix elements $H_{nm}$ and wavefunction $\psi(m)$ at $\chi = 50 $. 
    The number of sample $N_s$ is fixed at 2048.
    The dark-colored and light-colored bars represent the original and the semistochastic algorithm with $\epsilon = 0.01$ and $N_{\epsilon} = 100$.
    The speed-up ratio for wavefunction computation is indicated above the light-colored bars.
    \textbf{c}:
    Timing for local energy evaluations as a function of the chain lengths (from \ce{H16} to \ce{H48}). 
    The number of sample $N_s$ is fixed at $2048$, while the bond dimensions $\chi$ are set to $20$, $50$ and $100$.
    \textbf{d}:
    Energy convergence for variational optimization with different $\chi$ and $N_s = 2 \times 10^5$.
    The shaded regions indicates the chemistry accuracy (1 kcal/mol).
    }
    \label{fig:H50}
\end{figure}

Figure \ref{fig:H50}a illustrates the behaviour of the semistochastic algorithm for local energy with respect to the choices of $\epsilon$ and $N_{\epsilon}$ for \ce{H50}.
We find that the mean energy converges systematically towards the exact value (black) with decreasing variance as $N_{\epsilon}$ increases. Note that the results obtained with $N_{\epsilon}=0$ 
exhibit a large bias. To strike a balance between computational efficiency and accuracy, 
we will use $\epsilon = 0.01$ and $N_{\epsilon} = 100$ in optimizaiton.
With these settings, the computational cost is reduced to just $0.05\%$ of the original, yielding an approximately 2000-fold speedup. Furthermore, to analyze the computational scaling for the evaluation of local energy, we investigate the 1D hydrogen chains with three different sizes (\ce{H16}, \ce{H32}, and \ce{H48}). Figure \ref{fig:H50}b presents the detailed breakdown for the computational time, including both the computation of Hamiltonian matrix elements and wavefunctions. 
Notably, the computation of matrix elements requires significantly less time than the wavefunction computation. Compared to the original scheme, the semistochastic algorithm achieves a speedup ratio of 43, 418, and 1359 for the wavefunction calculation.
The result of the computational scaling is summarized in Fig. \ref{fig:H50}c for different chain lengths. The scaling of the exact local energy evaluation is $\mathcal{O}(K^5)$, as
the evaluation of the \bdgrnn{1} ansatz scales as $\mathcal{O}(\chi^2K)$ and the number of non-zero matrix elements is $\mathcal{O}(K^4)$. In contrast, the semistochastic evaluation achieves a significant reduction in scaling, which decreases from $\mathcal{O}(K^5)$ to approximately $\mathcal{O}(K^2)$. Note that the scaling continues to decrease as $\chi$ increases, 
which results from the fact that the proportion of wavefunction evaluation becomes more dominant with increasing $\chi$.

We now investigate the optimization of the \bdgrnn{1} ansatz as a function of the bond dimension $\chi$ using the \ce{H50} chain. Figure \ref{fig:H50}d shows the energy convergence throughout the optimization iterations. The \bdgrnn{1} ansatz demonstrates better accuracy compared to MPS at the same $\chi$. For example, \bdgrnn{1} at $\chi=30$ yields an energy error $0.6$ mHa, whereas MPS calculations require $\chi > 50$ to attain the same level of accuracy.
This is likely due to the increased variational degree of freedoms in the ansatz.
To further enhance the accuracy, we apply the correlation factor cos-RBM($\alpha = 2$) for $\chi=20$, with its parameters partially initialized from the result of the optimized \bdgrnn{1}.
This approach further reduces the energy error to $0.2$ mHa, whereas the energy error for \bdgrnn without the correlation factor is $2.6$ mHa. Note that the parameters of \bdgrnn{1} are 46100 ($\chi = 10$), 172100 ($\chi = 20$), and 378100 ($\chi = 30$), while \bdgrnn{1} ($\chi=20$) with $\cos\text{-RBM} (\alpha=2)$ contains 192300 parameters. Therefore, the improvement in accuracy is achieved with only 20200 additional parameters compared to the original \bdgrnn{1} ansatz.


\subsection{Iron--sulfur cluster \ce{[Fe2S2(SCH3)4]^{2-}}}
\begin{figure}[!htp]
    \centering
     \includegraphics[width=0.98\linewidth]{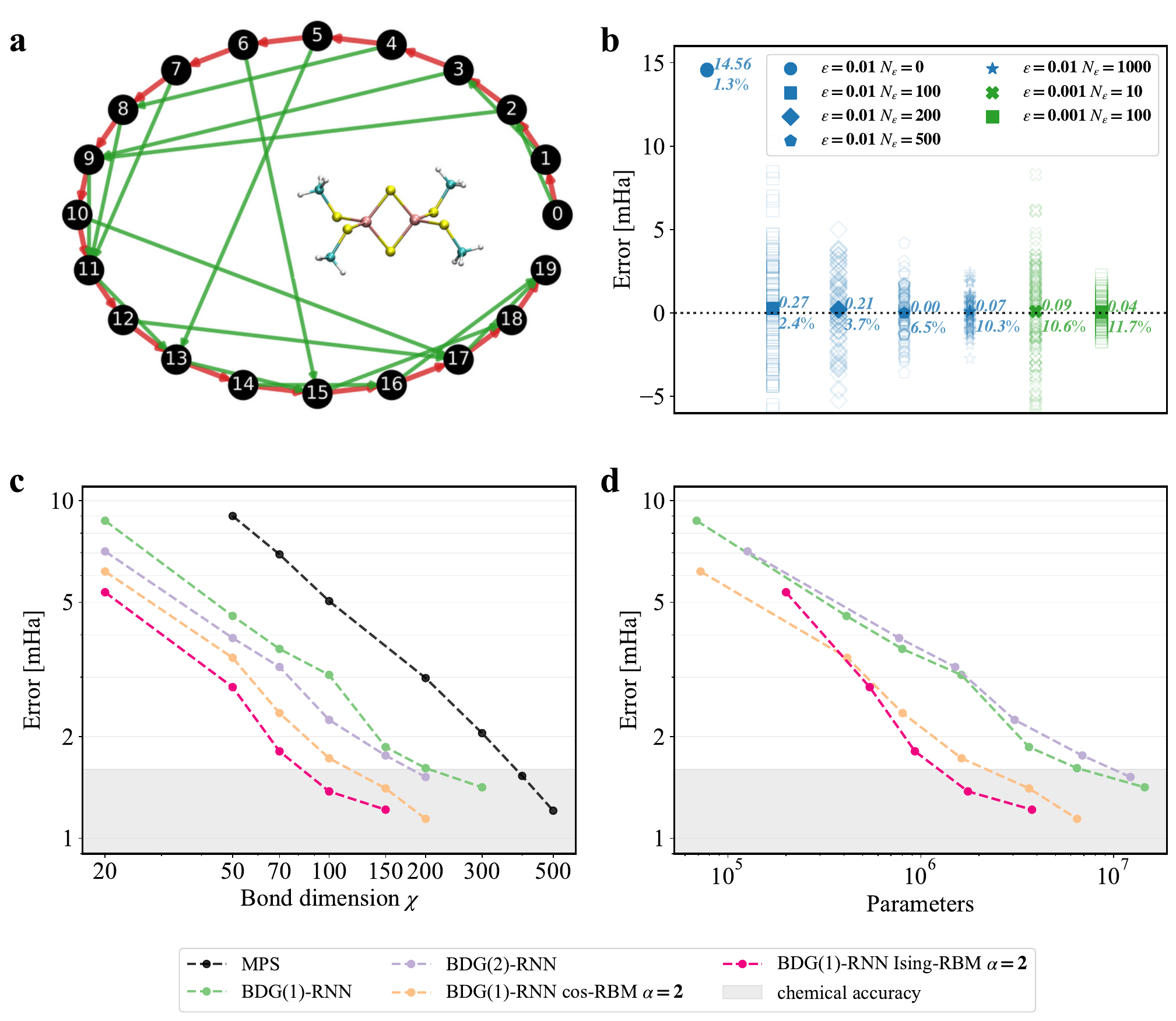}
    \caption{
    Results for the CAS(30e,2o) model of the iron--sulfur cluster \ce{[Fe2S2(SCH3)4]^{2-}}.
    \textbf{a}: 
    Illustration of the topology of \bdgrnn{1} and \bdgrnn{2}. The red arrows represent 
    \bdgrnn{1} (=MPS--RNN), while the green arrows
    represent the additional edges in \bdgrnn{2}. 
    \textbf{b}:
    Assessment of the accuracy, precision, and efficient the semistochastic algorithm for local energy with different hyperparameters $\epsilon$ and $N_{\epsilon}$ using the \bdgrnn{1} ansatz with $\chi=50$ and a sample size of $N_s = 10^7$. The error relative to the exact local energy is shown above each marker,
    and the percentage of computational cost relative to the original algorithm is shown below.
    \textbf{c},\textbf{d}:
    Energy convergence with respect to the bond dimension $\chi$ and the number of variational parameters for \bdgrnn with and without the cos-RBM and Ising-RBM correlators obtained using $N_s = 10^7$.
    The exact energy is taken from Ref. \cite{Li2017}.
    }
    \label{fig:Fe2S2}
\end{figure}

Due to the large number of $d$-orbital electrons and different kinds of spin and charge fluctuations\cite{Sharma2014, Li2017}, iron-sulfur clusters are typical examples for strongly correlated systems.
In this work, we selected the previously investigated \ce{[Fe2S2(SCH3)4]^{2-}} as the benchmark system with a CAS(30e, 20o) model, where the active space and integrals\cite{Li2017} were available
online\cite{linkToFCIDUMPfe2fe4}. To reduce orbital entanglement, we use the entanglement-minimized orbitals (EMOs) obtained using spin-adapted MPS with $\chi=100$\cite{Li2025EMO}.
Figure \ref{fig:Fe2S2}a illustrates the topological structures of the \bdgrnn{1} and \bdgrnn{2} architectures. In Fig. \ref{fig:Fe2S2}b, we compare the performance of the semistochastic algorithm for local energy using different $\epsilon$ and $N_{\epsilon}$ for the \bdgrnn{1} ansatz with $\chi=50$. Based on these results, we selected $\epsilon = 0.01$ and $N_{\epsilon} = 1000$
by balancing computational accuracy, precision, and efficiency, which reduces the computational cost by roughly an order of magnitude compared to the original scheme.

We present the energy convergence as a function of the bond dimension $\chi$ and the number of variational parameters in Figs. \ref{fig:Fe2S2}c and \ref{fig:Fe2S2}d, respectively.
Similarly to the hydrogen chain, \bdgrnn{1} shows a marked reduction in energy error compared to the standard MPS with identical bond dimensions $\chi$.
At $\chi = 200$, \bdgrnn{1} achieves chemical accuracy (1 kcal/mol), while standard MPS requires $\chi = 400$ to achieve a comparable accuracy. 
Compared to \bdgrnn{1}, \bdgrnn{2} demonstrates a slight improvement on the energy error at the same $\chi$. A large improvement is achieved by incorporating the cos-RBM correlator with the parameters $\alpha$ set to $2$. At $\chi = 100 $, the cos-RBM achieves an error of $1.7$ mHa (approaching chemical accuracy), outperforming the \bdgrnn{1} result ($3.0$ mHa).
Increasing the bond dimension to $\chi  = 200$ further reduces the cos-RBM error to $1.1$ mHa.
Finally, the more sophisticated correlator Ising-RBM with the parameter $\alpha = 2 $ yields additional gains in accuracy at the same $\chi$ and the number of parameters.
In particular, it reaches the chemical accuracy when $\chi = 100$. 


\subsection{Three-dimensional $3 \times 3 \times 2$ hydrogen cluster \ce{H18}}
\begin{figure}[!htp]
    \centering
    \includegraphics[width=0.80\linewidth]{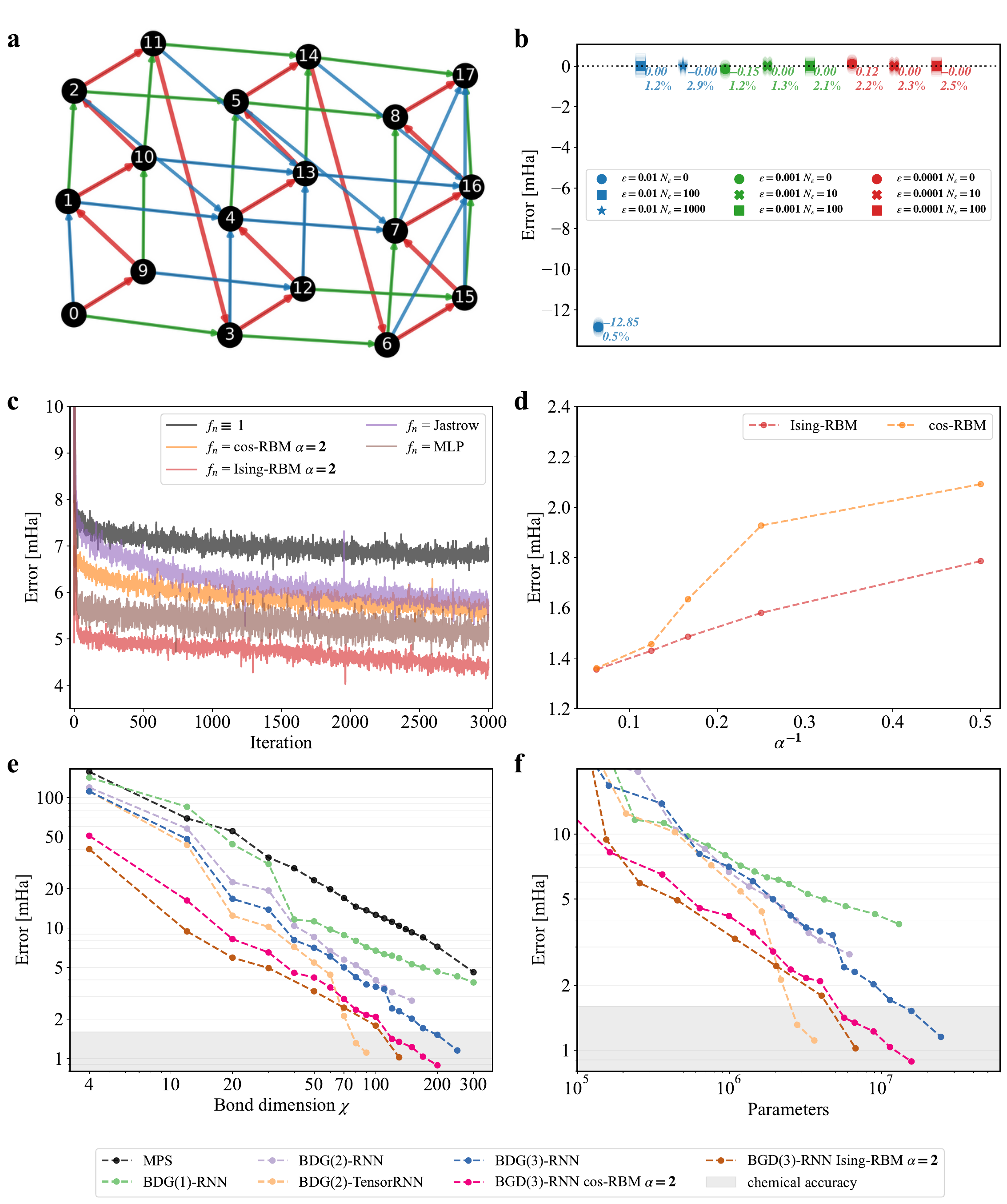}
    \captionsetup{skip=20pt} 
    \caption{
    Results for the three-dimensional $3\times 3\times 2$ hydrogen clusters \ce{H18}.
    \textbf{a}:
    Illustration of the topologies of the \bdgrnn variants. Red edges denote the base \bdgrnn{1} structure, green edges represent the additional edges in \bdgrnn{2}, while blue edges indicate
    the additional edges in \bdgrnn{3}.
    \textbf{b}:
    Assessment of the accuracy, precision, and efficient the semistochastic algorithm for local energy with different hyperparameters $\epsilon$ and $N_{\epsilon}$ using the \bdgrnn{3} ansatz with $\chi=100$ and a sample size of $N_s = 2 \times 10^5$.
    The error with respect to the exact local energy is shown above each marker.
    In later calculations, the values of $\epsilon$ and $N_{\epsilon}$ are set to 0.01 and 100, respectively.
    \textbf{c}:
    The optimization process for different correlators (cos-RBM, Ising-RBM, Jastrow and MLP) using
    \bdgrnn{3} with $\chi = 40$.
    The reference energy is taken from the MPS energy with $\chi = 2000$. 
    \textbf{d}:
    Absolute error of energy (mHa) as a function of $\alpha^{-1}$ using cos-RBM and
    Ising-RBM correlators on top of \bdgrnn{3} with $\chi=100$.
    \textbf{e}, \textbf{f}:
    Energy convergence with respect to the bond dimension $\chi$ and the number of variational parameters for \bdgrnn with and without the cos-RBM and Ising-RBM correlators obtained using $N_s = 10^6$.
    }
    \label{fig:H18}
\end{figure}

To examine the potential of the \bdgrnn ansatz and RBM-inspired correlator, 
we select a more challenging 3D system --- a $3 \times 3 \times 2$ hydrogen cluster \ce{H18}, 
where MPS require much larger bond dimensions $(\chi)$ to accurately capture the entanglement of the system. We set the interatomic distance to $4.0\Bohr$ between the nearest-neighbour 
hydrogen atoms. The OAOs obtained in the STO-3G basis are ordered with the Fiedler ordering\cite{barcza2011quantum,olivares2015ab} for the calculations using MPS and \bdgrnn{k} ($k\le 3$). The topologies of \bdgrnn are illustrated in Fig. \ref{fig:H18}a.
For local energy evaluation, we systematically tested the hyperparameters $\epsilon$ and $N_\epsilon$
in Fig. \ref{fig:H18}b. The choice of $\epsilon = 0.01$ and $N_{\epsilon} = 100$ achieves approximately an 80-fold speedup compared to the original algorithm, while 
maintaining energy errors below 0.1 mHa, indicating a good trade-off between computational efficiency and accuracy. 

In Fig. \ref{fig:H18}c, we evaluate the performance of correlators on top of \bdgrnn{3} with $\chi=40$, including cos-RBM, Ising-RBM, Jastrow ($f_{\rm{Jastrow}}(n) = \ee^{\sum_{ij}^{2K}W_{ij}n_in_j}$ with
$(W_{ij}) \in \mathbb{R}^{2K \times 2K}$), and a multilayer perceptron (MLP) defined by
\begin{equation}
    f_{\rm{mlplike}}(n) = \prod_{k=1}^{h_2}\cos\left(
    b_{1,k} + 2\uppi \sum_{j=1}^{h_1}W_{1, jk} \cdot f_0(b_{0, j} + \sum_i^{2K} W_{0, ij}n_i)
    \right),
\end{equation}
where $(W_{0,ij}) \in \mathbb{R}^{2K \times h_1}$ and $(W_{1,jk}) \in \mathbb{R}^{h_1 \times h_2}$ are real weights matrices ($h_1=4K$ and $h_2=4K$), while $(b_{0,j}) \in \mathbb{R}^{h_1}$ and $(b_{1,k}) \in \mathbb{R}^{h_2}$ are bias vectors, and the activation function is chosen as $f_0(z) = z^2 + z$.
All these correlation factors significantly improve the accuracy, with the Ising-RBM demonstrating the best performance among them. In addition, we investigated the effects of increasing $\alpha$ in cos-RBM and Ising-RBM in Fig. \ref{fig:H18}d. At $\alpha=2$, the Ising-RBM parameters are derived from further optimization of the cos-RBM. For $\alpha > 2$, parameters were updated utilizing the optimized parameters from the preceding $\alpha$ as initial conditions.
We observe that as $\alpha$ increases, the energy error further decreases slightly.
Notably, when $\alpha =16$, the energy errors of cos-RBM and Ising-RBM converge to nearly identical values.

We compare the convergence behaviors of MPS and \bdgrnn in Figs. \ref{fig:H18}e and \ref{fig:H18}f,
respectively. MPS demonstrates slow convergence toward the ground state energy with increasing bond dimension, requiring $\chi > 500$ to achieve the chemical accuracy.
The energy of \bdgrnn{1} also exhibits slow convergence, although its error is systematically smaller compared to that of the MPS. Compared with \bdgrnn{1}, both \bdgrnn{2} and \bdgrnn{3} show significantly faster convergence with the increasing bond dimension.
For example, the energy error of \bdgrnn{3}
approaches 1.1 mHa at $\chi = 250$, achieving the chemical accuracy.
Compared with \bdgrnn{2}, \bdgrnn{3} does not exhibit significant improvement over \bdgrnn{2},
which is likely due to the quasi-3D structure of the molecule.
We integrate the cos-RBM and Ising-RBM correlators with $\alpha=2$ into the \bdgrnn{3}.
With the correlation factors, the same accuracy as the previous \bdgrnn{3} at $\chi = 250$ can now be achieved at a reduced $\chi = 150$. We also examine the inclusion of the tensor terms into \bdgrnn{2}, denoted BDG(2)--TensorRNN. Notably, BDG(2)--TensorRNN achieves accuracy comparable to \bdgrnn{3} at $\chi = 250$ with a reduced bond dimension of $\chi = 100$, while the number of variational parameters for reaching the chemical accuracy is also significantly reduced.

\section{Conclusion}\label{sec:conlusion}
In this work, we propose a bounded--degree graph RNN (\bdgrnn) ans\"{a}tze,
which can represent complex entanglement structures 
in molecular systems using the RNN framework. 
\bdgrnn can be initialized using the MPS parameters, alleviating the optimization difficulties caused by random initialization parameters. To further improve the accuracy,
we introduce RBM-inspired correlators (cos-RBM and Ising-RBM variants) that significantly enhance NQS expressivity, while remaining full compatibility with the autoregressive sampling via reweighting technique. Additionally, we introduce a semistochastic scheme for local energy evaluation that significantly reduces computational costs without significantly affecting accuracy. These ans\"atze and techniques are implemented in the open-source \textsc{PyNQS}\cite{pynqs_github} package, which facilitates the exploration
of NQS methodologies. We validated our methodology on several prototypical
systems, including the 1D hydrogen chain \ce{H50}, the iron–sulfur cluster \ce{[Fe2S2(SCH3)4]^2-} with the CAS(30e, 20o) active space, 
and the more challenging 3D $3 \times 3 \times 2$ hydrogen cluster \ce{H18},
where the chemical accuracy is achieved across all systems using \bdgrnn.
In particular, a large systematic enhancement in accuracy is observed using RBM-inspired correlators.

While \bdgrnn demonstrates superior accuracy over MPS at the same bond dimensions, further developments are necessary in order to apply it to large-scale strongly correlated systems.
These include incorporating established techniques for using symmetry and mixed precision schemes,
which would substantially reduce both memory requirements and computational costs.
In addition, future work will also apply advanced optimization techniques such as minimum-step stochastic reconfiguration\cite{Chen2024a,Rende2024}, which could significantly accelerate the VMC optimization. 
We hope that the algorithms and software
presented in this work will stimulate future developments in
state-of-the-art NQS methods for quantum chemistry.

\begin{acknowledgement}
This work was supported by the Innovation Program for Quantum Science and Technology (Grant No. 2023ZD0300200) and the Fundamental Research Funds for the Central Universities.
\end{acknowledgement}

\begin{suppinfo}\label{supp}
Detailed algorithms, derivation of the energy gradient in the presence of correlators,
values of $B=\braket{|f_n|^2}$ for different molecules.
\end{suppinfo}

\bibliography{main}

\providecommand{\latin}[1]{#1}
\makeatletter
\providecommand{\doi}
  {\begingroup\let\do\@makeother\dospecials
  \catcode`\{=1 \catcode`\}=2 \doi@aux}
\providecommand{\doi@aux}[1]{\endgroup\texttt{#1}}
\makeatother
\providecommand*\mcitethebibliography{\thebibliography}
\csname @ifundefined\endcsname{endmcitethebibliography}
  {\let\endmcitethebibliography\endthebibliography}{}
\begin{mcitethebibliography}{61}
\providecommand*\natexlab[1]{#1}
\providecommand*\mciteSetBstSublistMode[1]{}
\providecommand*\mciteSetBstMaxWidthForm[2]{}
\providecommand*\mciteBstWouldAddEndPuncttrue
  {\def\EndOfBibitem{\unskip.}}
\providecommand*\mciteBstWouldAddEndPunctfalse
  {\let\EndOfBibitem\relax}
\providecommand*\mciteSetBstMidEndSepPunct[3]{}
\providecommand*\mciteSetBstSublistLabelBeginEnd[3]{}
\providecommand*\EndOfBibitem{}
\mciteSetBstSublistMode{f}
\mciteSetBstMaxWidthForm{subitem}{(\alph{mcitesubitemcount})}
\mciteSetBstSublistLabelBeginEnd
  {\mcitemaxwidthsubitemform\space}
  {\relax}
  {\relax}

\bibitem[Grimme(2003)]{Grimme2003_MP2}
Grimme,~S. Improved second-order M{\o}ller--Plesset perturbation theory by
  separate scaling of parallel-and antiparallel-spin pair correlation energies.
  \emph{J. Chem. Phys.} \textbf{2003}, \emph{118}, 9095--9102\relax
\mciteBstWouldAddEndPuncttrue
\mciteSetBstMidEndSepPunct{\mcitedefaultmidpunct}
{\mcitedefaultendpunct}{\mcitedefaultseppunct}\relax
\EndOfBibitem
\bibitem[Werner and Knowles(1988)Werner, and Knowles]{Werner1988_CI}
Werner,~H.-J.; Knowles,~P.~J. An efficient internally contracted
  multiconfiguration--reference configuration interaction method. \emph{J.
  Chem. Phys.} \textbf{1988}, \emph{89}, 5803--5814\relax
\mciteBstWouldAddEndPuncttrue
\mciteSetBstMidEndSepPunct{\mcitedefaultmidpunct}
{\mcitedefaultendpunct}{\mcitedefaultseppunct}\relax
\EndOfBibitem
\bibitem[Bartlett and Musia{\l}(2007)Bartlett, and Musia{\l}]{Bartlett2007_CC}
Bartlett,~R.~J.; Musia{\l},~M. Coupled-cluster theory in quantum chemistry.
  \emph{Rev. Mod. Phys.} \textbf{2007}, \emph{79}, 291--352\relax
\mciteBstWouldAddEndPuncttrue
\mciteSetBstMidEndSepPunct{\mcitedefaultmidpunct}
{\mcitedefaultendpunct}{\mcitedefaultseppunct}\relax
\EndOfBibitem
\bibitem[Kohn \latin{et~al.}(1996)Kohn, Becke, and Parr]{Kohn1996_DFT}
Kohn,~W.; Becke,~A.~D.; Parr,~R.~G. Density functional theory of electronic
  structure. \emph{J. Chem. Phys.} \textbf{1996}, \emph{100},
  12974--12980\relax
\mciteBstWouldAddEndPuncttrue
\mciteSetBstMidEndSepPunct{\mcitedefaultmidpunct}
{\mcitedefaultendpunct}{\mcitedefaultseppunct}\relax
\EndOfBibitem
\bibitem[Ceperley \latin{et~al.}(1977)Ceperley, Chester, and
  Kalos]{Ceperley1977_VMC}
Ceperley,~D.; Chester,~G.~V.; Kalos,~M.~H. Monte Carlo simulation of a
  many-fermion study. \emph{Phys. Rev. B} \textbf{1977}, \emph{16}, 3081\relax
\mciteBstWouldAddEndPuncttrue
\mciteSetBstMidEndSepPunct{\mcitedefaultmidpunct}
{\mcitedefaultendpunct}{\mcitedefaultseppunct}\relax
\EndOfBibitem
\bibitem[Toulouse \latin{et~al.}(2016)Toulouse, Assaraf, and
  Umrigar]{toulouse2016introduction}
Toulouse,~J.; Assaraf,~R.; Umrigar,~C.~J. \emph{Adv. Quantum Chem.}; Elsevier,
  2016; Vol.~73; pp 285--314\relax
\mciteBstWouldAddEndPuncttrue
\mciteSetBstMidEndSepPunct{\mcitedefaultmidpunct}
{\mcitedefaultendpunct}{\mcitedefaultseppunct}\relax
\EndOfBibitem
\bibitem[Austin \latin{et~al.}(2012)Austin, Zubarev, and
  Lester~Jr]{austin2012quantum}
Austin,~B.~M.; Zubarev,~D.~Y.; Lester~Jr,~W.~A. Quantum Monte Carlo and related
  approaches. \emph{Chem. Rev.} \textbf{2012}, \emph{112}, 263--288\relax
\mciteBstWouldAddEndPuncttrue
\mciteSetBstMidEndSepPunct{\mcitedefaultmidpunct}
{\mcitedefaultendpunct}{\mcitedefaultseppunct}\relax
\EndOfBibitem
\bibitem[Kalos(1962)]{Kalos1962_GFMC}
Kalos,~M.~H. Monte Carlo calculations of the ground state of three-and
  four-body nuclei. \emph{Phys. Rev.} \textbf{1962}, \emph{128}, 1791\relax
\mciteBstWouldAddEndPuncttrue
\mciteSetBstMidEndSepPunct{\mcitedefaultmidpunct}
{\mcitedefaultendpunct}{\mcitedefaultseppunct}\relax
\EndOfBibitem
\bibitem[Sorella(1998)]{Sorella1998_GFMC}
Sorella,~S. Green function Monte Carlo with stochastic reconfiguration.
  \emph{Phys. Rev. Lett.} \textbf{1998}, \emph{80}, 4558\relax
\mciteBstWouldAddEndPuncttrue
\mciteSetBstMidEndSepPunct{\mcitedefaultmidpunct}
{\mcitedefaultendpunct}{\mcitedefaultseppunct}\relax
\EndOfBibitem
\bibitem[Motta and Zhang(2018)Motta, and Zhang]{motta2018ab}
Motta,~M.; Zhang,~S. Ab initio computations of molecular systems by the
  auxiliary-field quantum Monte Carlo method. \emph{WIREs Comput. Mol. Sci.}
  \textbf{2018}, \emph{8}, e1364\relax
\mciteBstWouldAddEndPuncttrue
\mciteSetBstMidEndSepPunct{\mcitedefaultmidpunct}
{\mcitedefaultendpunct}{\mcitedefaultseppunct}\relax
\EndOfBibitem
\bibitem[Lee \latin{et~al.}(2022)Lee, Pham, and Reichman]{lee2022twenty}
Lee,~J.; Pham,~H.~Q.; Reichman,~D.~R. Twenty years of auxiliary-field quantum
  Monte Carlo in quantum chemistry: An overview and assessment on main group
  chemistry and bond-breaking. \emph{J. Chem. Theory Comput.} \textbf{2022},
  \emph{18}, 7024--7042\relax
\mciteBstWouldAddEndPuncttrue
\mciteSetBstMidEndSepPunct{\mcitedefaultmidpunct}
{\mcitedefaultendpunct}{\mcitedefaultseppunct}\relax
\EndOfBibitem
\bibitem[Carleo and Troyer(2017)Carleo, and Troyer]{Carleo2017}
Carleo,~G.; Troyer,~M. Solving the quantum many-body problem with artificial
  neural networks. \emph{Science} \textbf{2017}, \emph{355}, 602--606\relax
\mciteBstWouldAddEndPuncttrue
\mciteSetBstMidEndSepPunct{\mcitedefaultmidpunct}
{\mcitedefaultendpunct}{\mcitedefaultseppunct}\relax
\EndOfBibitem
\bibitem[Hermann \latin{et~al.}(2023)Hermann, Spencer, Choo, Mezzacapo,
  Foulkes, Pfau, Carleo, and Noé]{Hermann2023_reviews}
Hermann,~J.; Spencer,~J.; Choo,~K.; Mezzacapo,~A.; Foulkes,~W. M.~C.; Pfau,~D.;
  Carleo,~G.; Noé,~F. Ab initio quantum chemistry with neural-network
  wavefunctions. \emph{Nat. Rev. Chem.} \textbf{2023}, \emph{7}, 692--709\relax
\mciteBstWouldAddEndPuncttrue
\mciteSetBstMidEndSepPunct{\mcitedefaultmidpunct}
{\mcitedefaultendpunct}{\mcitedefaultseppunct}\relax
\EndOfBibitem
\bibitem[Hibat-Allah \latin{et~al.}(2020)Hibat-Allah, Ganahl, Hayward, Melko,
  and Carrasquilla]{hibat-allah_recurrent_2020}
Hibat-Allah,~M.; Ganahl,~M.; Hayward,~L.~E.; Melko,~R.~G.; Carrasquilla,~J.
  Recurrent neural network wave functions. \emph{Phys. Rev. Res.}
  \textbf{2020}, \emph{2}, 023358\relax
\mciteBstWouldAddEndPuncttrue
\mciteSetBstMidEndSepPunct{\mcitedefaultmidpunct}
{\mcitedefaultendpunct}{\mcitedefaultseppunct}\relax
\EndOfBibitem
\bibitem[Wu \latin{et~al.}(2023)Wu, Rossi, Vicentini, and
  Carleo]{wu_tensor-network_2023}
Wu,~D.; Rossi,~R.; Vicentini,~F.; Carleo,~G. From tensor-network quantum states
  to tensorial recurrent neural networks. \emph{Phys. Rev. Res.} \textbf{2023},
  \emph{5}, L032001\relax
\mciteBstWouldAddEndPuncttrue
\mciteSetBstMidEndSepPunct{\mcitedefaultmidpunct}
{\mcitedefaultendpunct}{\mcitedefaultseppunct}\relax
\EndOfBibitem
\bibitem[Choo \latin{et~al.}(2019)Choo, Neupert, and Carleo]{Choo2019_CNNS}
Choo,~K.; Neupert,~T.; Carleo,~G. Two-dimensional frustrated J1-J2 model
  studied with neural network quantum states. \emph{Phys. Rev. B}
  \textbf{2019}, \emph{100}, 125124\relax
\mciteBstWouldAddEndPuncttrue
\mciteSetBstMidEndSepPunct{\mcitedefaultmidpunct}
{\mcitedefaultendpunct}{\mcitedefaultseppunct}\relax
\EndOfBibitem
\bibitem[Barrett \latin{et~al.}(2022)Barrett, Malyshev, and
  Lvovsky]{Barrett2022}
Barrett,~T.~D.; Malyshev,~A.; Lvovsky,~A. Autoregressive neural-network
  wavefunctions for ab initio quantum chemistry. \emph{Nat. Mac. Intell.}
  \textbf{2022}, \emph{4}, 351--358\relax
\mciteBstWouldAddEndPuncttrue
\mciteSetBstMidEndSepPunct{\mcitedefaultmidpunct}
{\mcitedefaultendpunct}{\mcitedefaultseppunct}\relax
\EndOfBibitem
\bibitem[Chen \latin{et~al.}(2023)Chen, Newhouse, Chen, Luo, and
  Soljacic]{chen2023antn}
Chen,~Z.; Newhouse,~L.; Chen,~E.; Luo,~D.; Soljacic,~M. Antn: Bridging
  autoregressive neural networks and tensor networks for quantum many-body
  simulation. \emph{Adv. Neural Inf. Process. Syst.} \textbf{2023}, \emph{36},
  450--476\relax
\mciteBstWouldAddEndPuncttrue
\mciteSetBstMidEndSepPunct{\mcitedefaultmidpunct}
{\mcitedefaultendpunct}{\mcitedefaultseppunct}\relax
\EndOfBibitem
\bibitem[Zhang and Di~Ventra(2023)Zhang, and Di~Ventra]{Zhang2023}
Zhang,~Y.-H.; Di~Ventra,~M. Transformer quantum state: A multipurpose model for
  quantum many-body problems. \emph{Phys. Rev. B} \textbf{2023}, \emph{107},
  075147\relax
\mciteBstWouldAddEndPuncttrue
\mciteSetBstMidEndSepPunct{\mcitedefaultmidpunct}
{\mcitedefaultendpunct}{\mcitedefaultseppunct}\relax
\EndOfBibitem
\bibitem[Wu \latin{et~al.}(2023)Wu, Guo, Fan, Zhou, and Shang]{Wu2023nnqs}
Wu,~Y.; Guo,~C.; Fan,~Y.; Zhou,~P.; Shang,~H. NNQS-transformer: an efficient
  and scalable neural network quantum states approach for ab initio quantum
  chemistry. Proceedings of the International Conference for High Performance
  Computing, Networking, Storage and Analysis. 2023; pp 1--13\relax
\mciteBstWouldAddEndPuncttrue
\mciteSetBstMidEndSepPunct{\mcitedefaultmidpunct}
{\mcitedefaultendpunct}{\mcitedefaultseppunct}\relax
\EndOfBibitem
\bibitem[Rende \latin{et~al.}(2024)Rende, Viteritti, Bardone, Becca, and
  Goldt]{Rende2024}
Rende,~R.; Viteritti,~L.~L.; Bardone,~L.; Becca,~F.; Goldt,~S. A simple linear
  algebra identity to optimize Large-Scale Neural Network Quantum States.
  \emph{Commun. Phys.} \textbf{2024}, \emph{7}\relax
\mciteBstWouldAddEndPuncttrue
\mciteSetBstMidEndSepPunct{\mcitedefaultmidpunct}
{\mcitedefaultendpunct}{\mcitedefaultseppunct}\relax
\EndOfBibitem
\bibitem[Knitter \latin{et~al.}(2025)Knitter, Zhao, Stokes, Ganahl,
  Leichenauer, and Veerapaneni]{knitter2025retentive}
Knitter,~O.; Zhao,~D.; Stokes,~J.; Ganahl,~M.; Leichenauer,~S.; Veerapaneni,~S.
  Retentive neural quantum states: efficient ans{\"a}tze for ab initio quantum
  chemistry. \emph{Mach. Learn.: Sci. Technol.} \textbf{2025}, \emph{6},
  025022\relax
\mciteBstWouldAddEndPuncttrue
\mciteSetBstMidEndSepPunct{\mcitedefaultmidpunct}
{\mcitedefaultendpunct}{\mcitedefaultseppunct}\relax
\EndOfBibitem
\bibitem[Liu and Clark(2024)Liu, and Clark]{Liu2024}
Liu,~A.-J.; Clark,~B.~K. Neural network backflow for ab-initio quantum
  chemistry. \emph{Phys. Rev. B} \textbf{2024}, \emph{110}, 115137\relax
\mciteBstWouldAddEndPuncttrue
\mciteSetBstMidEndSepPunct{\mcitedefaultmidpunct}
{\mcitedefaultendpunct}{\mcitedefaultseppunct}\relax
\EndOfBibitem
\bibitem[Kan \latin{et~al.}(2025)Kan, Tian, Wu, Zhang, and Shang]{Kan2025}
Kan,~B.; Tian,~Y.; Wu,~Y.; Zhang,~Y.; Shang,~H. Bridging the Gap between
  Transformer-Based Neural Networks and Tensor Networks for Quantum Chemistry.
  \emph{J. Chem. Theory Comput.} \textbf{2025}, \emph{21}, 3426--3439\relax
\mciteBstWouldAddEndPuncttrue
\mciteSetBstMidEndSepPunct{\mcitedefaultmidpunct}
{\mcitedefaultendpunct}{\mcitedefaultseppunct}\relax
\EndOfBibitem
\bibitem[Du and Chan(2025)Du, and Chan]{du2025neuralized}
Du,~S.-J.; Chan,~G.~K. Neuralized Fermionic Tensor Networks for Quantum
  Many-Body Systems. \emph{arXiv preprint arXiv:2506.08329} \textbf{2025},
  \relax
\mciteBstWouldAddEndPunctfalse
\mciteSetBstMidEndSepPunct{\mcitedefaultmidpunct}
{}{\mcitedefaultseppunct}\relax
\EndOfBibitem
\bibitem[Li \latin{et~al.}(2023)Li, Huang, Zhang, Li, Cao, Lv, and
  Hu]{xiang2023}
Li,~X.; Huang,~J.-C.; Zhang,~G.-Z.; Li,~H.-E.; Cao,~C.-S.; Lv,~D.; Hu,~H.-S.
  Non-stochastic Optimization Algorithm for Neural-network Quantum States.
  \emph{J. Chem. Theory Comput.} \textbf{2023}, \emph{19}, 8156--8165\relax
\mciteBstWouldAddEndPuncttrue
\mciteSetBstMidEndSepPunct{\mcitedefaultmidpunct}
{\mcitedefaultendpunct}{\mcitedefaultseppunct}\relax
\EndOfBibitem
\bibitem[Li \latin{et~al.}(2024)Li, Huang, Zhang, Li, Shen, Zhao, and
  Hu]{xiang2024}
Li,~X.; Huang,~J.-C.; Zhang,~G.-Z.; Li,~H.-E.; Shen,~Z.-P.; Zhao,~C.; Hu,~H.-S.
  Improved Optimization for the Neural-network Quantum States and Tests on the
  Chromium Dimer. \emph{J. Chem. Phys.} \textbf{2024}, \emph{160}\relax
\mciteBstWouldAddEndPuncttrue
\mciteSetBstMidEndSepPunct{\mcitedefaultmidpunct}
{\mcitedefaultendpunct}{\mcitedefaultseppunct}\relax
\EndOfBibitem
\bibitem[Liu and Clark(2025)Liu, and Clark]{Liu2025}
Liu,~A.-J.; Clark,~B.~K. Efficient optimization of neural network backflow for
  ab-initio quantum chemistry. \emph{arXiv preprint arXiv:2502.18843}
  \textbf{2025}, \relax
\mciteBstWouldAddEndPunctfalse
\mciteSetBstMidEndSepPunct{\mcitedefaultmidpunct}
{}{\mcitedefaultseppunct}\relax
\EndOfBibitem
\bibitem[Zhao \latin{et~al.}(2023)Zhao, Stokes, and Veerapaneni]{Zhao2023}
Zhao,~T.; Stokes,~J.; Veerapaneni,~S. Scalable neural quantum states
  architecture for quantum chemistry. \emph{Mach. Learn.: Sci. Technol.}
  \textbf{2023}, \emph{4}, 025034\relax
\mciteBstWouldAddEndPuncttrue
\mciteSetBstMidEndSepPunct{\mcitedefaultmidpunct}
{\mcitedefaultendpunct}{\mcitedefaultseppunct}\relax
\EndOfBibitem
\bibitem[Schwarz \latin{et~al.}(2017)Schwarz, Alavi, and Booth]{Schwarz2017}
Schwarz,~L.~R.; Alavi,~A.; Booth,~G.~H. Projector Quantum Monte Carlo Method
  for Nonlinear Wave Functions. \emph{Phys. Rev. Lett.} \textbf{2017},
  \emph{118}, 176403\relax
\mciteBstWouldAddEndPuncttrue
\mciteSetBstMidEndSepPunct{\mcitedefaultmidpunct}
{\mcitedefaultendpunct}{\mcitedefaultseppunct}\relax
\EndOfBibitem
\bibitem[Wei and Neuscamman(2018)Wei, and Neuscamman]{Wei2018}
Wei,~H.; Neuscamman,~E. Reduced scaling Hilbert space variational Monte Carlo.
  \emph{J. Chem. Phys.} \textbf{2018}, \emph{149}\relax
\mciteBstWouldAddEndPuncttrue
\mciteSetBstMidEndSepPunct{\mcitedefaultmidpunct}
{\mcitedefaultendpunct}{\mcitedefaultseppunct}\relax
\EndOfBibitem
\bibitem[Sabzevari and Sharma(2018)Sabzevari, and Sharma]{Sabzevari2018}
Sabzevari,~I.; Sharma,~S. Improved speed and scaling in orbital space
  variational Monte Carlo. \emph{J. Chem. Theory Comput.} \textbf{2018},
  \emph{14}, 6276--6286\relax
\mciteBstWouldAddEndPuncttrue
\mciteSetBstMidEndSepPunct{\mcitedefaultmidpunct}
{\mcitedefaultendpunct}{\mcitedefaultseppunct}\relax
\EndOfBibitem
\bibitem[Levin and Peres(2017)Levin, and Peres]{Levin2017_Markov}
Levin,~D.~A.; Peres,~Y. \emph{Markov chains and mixing times}; American
  Mathematical Soc., 2017; Vol. 107\relax
\mciteBstWouldAddEndPuncttrue
\mciteSetBstMidEndSepPunct{\mcitedefaultmidpunct}
{\mcitedefaultendpunct}{\mcitedefaultseppunct}\relax
\EndOfBibitem
\bibitem[Griewank and Walther(2008)Griewank, and Walther]{Griewank2008_AD}
Griewank,~A.; Walther,~A. \emph{Evaluating derivatives: principles and
  techniques of algorithmic differentiation}; SIAM, 2008\relax
\mciteBstWouldAddEndPuncttrue
\mciteSetBstMidEndSepPunct{\mcitedefaultmidpunct}
{\mcitedefaultendpunct}{\mcitedefaultseppunct}\relax
\EndOfBibitem
\bibitem[Robbins and Monro(1951)Robbins, and Monro]{robbins1951stochastic}
Robbins,~H.; Monro,~S. A stochastic approximation method. \emph{Ann. Math.
  Statist.} \textbf{1951}, 400--407\relax
\mciteBstWouldAddEndPuncttrue
\mciteSetBstMidEndSepPunct{\mcitedefaultmidpunct}
{\mcitedefaultendpunct}{\mcitedefaultseppunct}\relax
\EndOfBibitem
\bibitem[Kingma and Ba(2015)Kingma, and Ba]{kingma2014adam}
Kingma,~D.~P.; Ba,~J. Adam: A method for stochastic optimization. International
  Conference on Learning Representations. 2015\relax
\mciteBstWouldAddEndPuncttrue
\mciteSetBstMidEndSepPunct{\mcitedefaultmidpunct}
{\mcitedefaultendpunct}{\mcitedefaultseppunct}\relax
\EndOfBibitem
\bibitem[Loshchilov and Hutter(2019)Loshchilov, and
  Hutter]{loshchilov2019fixing}
Loshchilov,~I.; Hutter,~F. Fixing Weight Decay Regularization in Adam.
  International Conference on Learning Representations. 2019\relax
\mciteBstWouldAddEndPuncttrue
\mciteSetBstMidEndSepPunct{\mcitedefaultmidpunct}
{\mcitedefaultendpunct}{\mcitedefaultseppunct}\relax
\EndOfBibitem
\bibitem[Nomura \latin{et~al.}(2017)Nomura, Darmawan, Yamaji, and
  Imada]{Nomura2017}
Nomura,~Y.; Darmawan,~A.~S.; Yamaji,~Y.; Imada,~M. Restricted Boltzmann machine
  learning for solving strongly correlated quantum systems. \emph{Phys. Rev. B}
  \textbf{2017}, \emph{96}, 205152\relax
\mciteBstWouldAddEndPuncttrue
\mciteSetBstMidEndSepPunct{\mcitedefaultmidpunct}
{\mcitedefaultendpunct}{\mcitedefaultseppunct}\relax
\EndOfBibitem
\bibitem[Choo \latin{et~al.}(2020)Choo, Mezzacapo, and Carleo]{Choo2020}
Choo,~K.; Mezzacapo,~A.; Carleo,~G. Fermionic neural-network states for
  ab-initio electronic structure. \emph{Nat. Commun.} \textbf{2020}, \emph{11},
  1--7\relax
\mciteBstWouldAddEndPuncttrue
\mciteSetBstMidEndSepPunct{\mcitedefaultmidpunct}
{\mcitedefaultendpunct}{\mcitedefaultseppunct}\relax
\EndOfBibitem
\bibitem[Hibat-Allah \latin{et~al.}(2020)Hibat-Allah, Ganahl, Hayward, Melko,
  and Carrasquilla]{HibatAllah2020}
Hibat-Allah,~M.; Ganahl,~M.; Hayward,~L.~E.; Melko,~R.~G.; Carrasquilla,~J.
  Recurrent neural network wave functions. \emph{Phys. Rev. Res.}
  \textbf{2020}, \emph{2}, 023358\relax
\mciteBstWouldAddEndPuncttrue
\mciteSetBstMidEndSepPunct{\mcitedefaultmidpunct}
{\mcitedefaultendpunct}{\mcitedefaultseppunct}\relax
\EndOfBibitem
\bibitem[Chung \latin{et~al.}(2014)Chung, Gulcehre, Cho, and Bengio]{Chung2014}
Chung,~J.; Gulcehre,~C.; Cho,~K.; Bengio,~Y. Empirical evaluation of gated
  recurrent neural networks on sequence modeling. \emph{arXiv preprint
  arXiv:1412.3555} \textbf{2014}, \relax
\mciteBstWouldAddEndPunctfalse
\mciteSetBstMidEndSepPunct{\mcitedefaultmidpunct}
{}{\mcitedefaultseppunct}\relax
\EndOfBibitem
\bibitem[Hochreiter and Schmidhuber(1997)Hochreiter, and
  Schmidhuber]{Hochreiter1997}
Hochreiter,~S.; Schmidhuber,~J. Long short-term memory. \emph{Neural Comput.}
  \textbf{1997}, \emph{9}, 1735--1780\relax
\mciteBstWouldAddEndPuncttrue
\mciteSetBstMidEndSepPunct{\mcitedefaultmidpunct}
{\mcitedefaultendpunct}{\mcitedefaultseppunct}\relax
\EndOfBibitem
\bibitem[Barcza \latin{et~al.}(2011)Barcza, Legeza, Marti, and
  Reiher]{barcza2011quantum}
Barcza,~G.; Legeza,~{\"O}.; Marti,~K.~H.; Reiher,~M. Quantum-information
  analysis of electronic states of different molecular structures. \emph{Phys.
  Rev. A} \textbf{2011}, \emph{83}, 012508\relax
\mciteBstWouldAddEndPuncttrue
\mciteSetBstMidEndSepPunct{\mcitedefaultmidpunct}
{\mcitedefaultendpunct}{\mcitedefaultseppunct}\relax
\EndOfBibitem
\bibitem[Olivares-Amaya \latin{et~al.}(2015)Olivares-Amaya, Hu, Nakatani,
  Sharma, Yang, and Chan]{olivares2015ab}
Olivares-Amaya,~R.; Hu,~W.; Nakatani,~N.; Sharma,~S.; Yang,~J.; Chan,~G.~K. The
  ab-initio density matrix renormalization group in practice. \emph{J. Chem.
  Phys.} \textbf{2015}, \emph{142}\relax
\mciteBstWouldAddEndPuncttrue
\mciteSetBstMidEndSepPunct{\mcitedefaultmidpunct}
{\mcitedefaultendpunct}{\mcitedefaultseppunct}\relax
\EndOfBibitem
\bibitem[Tucker(1966)]{tucker1966some}
Tucker,~L.~R. Some mathematical notes on three-mode factor analysis.
  \emph{Psychometrika} \textbf{1966}, \emph{31}, 279--311\relax
\mciteBstWouldAddEndPuncttrue
\mciteSetBstMidEndSepPunct{\mcitedefaultmidpunct}
{\mcitedefaultendpunct}{\mcitedefaultseppunct}\relax
\EndOfBibitem
\bibitem[Deng \latin{et~al.}(2017)Deng, Li, and Das~Sarma]{deng2017quantum}
Deng,~D.-L.; Li,~X.; Das~Sarma,~S. Quantum entanglement in neural network
  states. \emph{Phys. Rev. X} \textbf{2017}, \emph{7}, 021021\relax
\mciteBstWouldAddEndPuncttrue
\mciteSetBstMidEndSepPunct{\mcitedefaultmidpunct}
{\mcitedefaultendpunct}{\mcitedefaultseppunct}\relax
\EndOfBibitem
\bibitem[Yang \latin{et~al.}(2020)Yang, Sugiyama, Tsuda, and
  Yanai]{Yanai2020_RBM}
Yang,~P.-J.; Sugiyama,~M.; Tsuda,~K.; Yanai,~T. Artificial Neural Networks
  Applied as Molecular Wave Function Solvers. \emph{J. Chem. Theory Comput.}
  \textbf{2020}, \emph{16}, 3513--3529\relax
\mciteBstWouldAddEndPuncttrue
\mciteSetBstMidEndSepPunct{\mcitedefaultmidpunct}
{\mcitedefaultendpunct}{\mcitedefaultseppunct}\relax
\EndOfBibitem
\bibitem[Becca and Sorella(2017)Becca, and Sorella]{becca2017quantum}
Becca,~F.; Sorella,~S. \emph{Quantum Monte Carlo approaches for correlated
  systems}; Cambridge University Press, 2017\relax
\mciteBstWouldAddEndPuncttrue
\mciteSetBstMidEndSepPunct{\mcitedefaultmidpunct}
{\mcitedefaultendpunct}{\mcitedefaultseppunct}\relax
\EndOfBibitem
\bibitem[pyn()]{pynqs_github}
\url{https://github.com/Quantum-Chemistry-Group-BNU/PyNQS}\relax
\mciteBstWouldAddEndPuncttrue
\mciteSetBstMidEndSepPunct{\mcitedefaultmidpunct}
{\mcitedefaultendpunct}{\mcitedefaultseppunct}\relax
\EndOfBibitem
\bibitem[Paszke \latin{et~al.}(2019)Paszke, Gross, Massa, Lerer, Bradbury,
  Chanan, Killeen, Lin, Gimelshein, Antiga, \latin{et~al.}
  others]{paszke2019pytorch}
Paszke,~A.; Gross,~S.; Massa,~F.; Lerer,~A.; Bradbury,~J.; Chanan,~G.;
  Killeen,~T.; Lin,~Z.; Gimelshein,~N.; Antiga,~L., \latin{et~al.}  Pytorch: An
  imperative style, high-performance deep learning library. \emph{Adv. Neural
  Inf. Process. Syst.} \textbf{2019}, \emph{32}\relax
\mciteBstWouldAddEndPuncttrue
\mciteSetBstMidEndSepPunct{\mcitedefaultmidpunct}
{\mcitedefaultendpunct}{\mcitedefaultseppunct}\relax
\EndOfBibitem
\bibitem[Li(2021)]{Li2021a}
Li,~Z. Expressibility of comb tensor network states (CTNS) for the P-cluster
  and the FeMo-cofactor of nitrogenase. \emph{Electron. Struct.} \textbf{2021},
  \emph{3}, 014001\relax
\mciteBstWouldAddEndPuncttrue
\mciteSetBstMidEndSepPunct{\mcitedefaultmidpunct}
{\mcitedefaultendpunct}{\mcitedefaultseppunct}\relax
\EndOfBibitem
\bibitem[Xiang \latin{et~al.}(2024)Xiang, Jia, Fang, and
  Li]{xiang2024distributed}
Xiang,~C.; Jia,~W.; Fang,~W.-H.; Li,~Z. Distributed Multi-GPU Ab Initio Density
  Matrix Renormalization Group Algorithm with Applications to the P-Cluster of
  Nitrogenase. \emph{J. Chem. Theory Comput.} \textbf{2024}, \emph{20},
  775--786\relax
\mciteBstWouldAddEndPuncttrue
\mciteSetBstMidEndSepPunct{\mcitedefaultmidpunct}
{\mcitedefaultendpunct}{\mcitedefaultseppunct}\relax
\EndOfBibitem
\bibitem[Sun \latin{et~al.}(2020)Sun, Zhang, Banerjee, Bao, Barbry, Blunt,
  Bogdanov, Booth, Chen, Cui, \latin{et~al.} others]{sun2020recent}
Sun,~Q.; Zhang,~X.; Banerjee,~S.; Bao,~P.; Barbry,~M.; Blunt,~N.~S.;
  Bogdanov,~N.~A.; Booth,~G.~H.; Chen,~J.; Cui,~Z.-H., \latin{et~al.}  Recent
  developments in the PySCF program package. \emph{J. Chem. Phys.}
  \textbf{2020}, \emph{153}\relax
\mciteBstWouldAddEndPuncttrue
\mciteSetBstMidEndSepPunct{\mcitedefaultmidpunct}
{\mcitedefaultendpunct}{\mcitedefaultseppunct}\relax
\EndOfBibitem
\bibitem[Schollw{\"o}ck(2011)]{Schollwoeck2011}
Schollw{\"o}ck,~U. The density-matrix renormalization group in the age of
  matrix product states. \emph{Ann. Phys.} \textbf{2011}, \emph{326},
  96--192\relax
\mciteBstWouldAddEndPuncttrue
\mciteSetBstMidEndSepPunct{\mcitedefaultmidpunct}
{\mcitedefaultendpunct}{\mcitedefaultseppunct}\relax
\EndOfBibitem
\bibitem[Hachmann \latin{et~al.}(2006)Hachmann, Cardoen, and
  Chan]{Hachmann2006}
Hachmann,~J.; Cardoen,~W.; Chan,~G.~K. Multireference correlation in long
  molecules with the quadratic scaling density matrix renormalization group.
  \emph{J. Chem. Phys.} \textbf{2006}, \emph{125}\relax
\mciteBstWouldAddEndPuncttrue
\mciteSetBstMidEndSepPunct{\mcitedefaultmidpunct}
{\mcitedefaultendpunct}{\mcitedefaultseppunct}\relax
\EndOfBibitem
\bibitem[Sharma and Chan(2014)Sharma, and Chan]{Sharma2014}
Sharma,~S.; Chan,~G.~K. Communication: A flexible multi-reference perturbation
  theory by minimizing the Hylleraas functional with matrix product states.
  \emph{J. Chem. Phys.} \textbf{2014}, \emph{141}\relax
\mciteBstWouldAddEndPuncttrue
\mciteSetBstMidEndSepPunct{\mcitedefaultmidpunct}
{\mcitedefaultendpunct}{\mcitedefaultseppunct}\relax
\EndOfBibitem
\bibitem[Li and Chan(2017)Li, and Chan]{Li2017}
Li,~Z.; Chan,~G. K.-L. Spin-projected matrix product states: Versatile tool for
  strongly correlated systems. \emph{J. Chem. Theory Comput.} \textbf{2017},
  \emph{13}, 2681--2695\relax
\mciteBstWouldAddEndPuncttrue
\mciteSetBstMidEndSepPunct{\mcitedefaultmidpunct}
{\mcitedefaultendpunct}{\mcitedefaultseppunct}\relax
\EndOfBibitem
\bibitem[lin(2017)]{linkToFCIDUMPfe2fe4}
\url{http://github.com/zhendongli2008/Active-space-model-for-Iron-Sulfur-Clusters},
  2017\relax
\mciteBstWouldAddEndPuncttrue
\mciteSetBstMidEndSepPunct{\mcitedefaultmidpunct}
{\mcitedefaultendpunct}{\mcitedefaultseppunct}\relax
\EndOfBibitem
\bibitem[Li(2025)]{Li2025EMO}
Li,~Z. Entanglement-minimized orbitals enable faster quantum simulation of
  molecules. \emph{arXiv preprint arXiv:2506.13386} \textbf{2025}, \relax
\mciteBstWouldAddEndPunctfalse
\mciteSetBstMidEndSepPunct{\mcitedefaultmidpunct}
{}{\mcitedefaultseppunct}\relax
\EndOfBibitem
\bibitem[Chen and Heyl(2024)Chen, and Heyl]{Chen2024a}
Chen,~A.; Heyl,~M. Efficient optimization of deep neural quantum states toward
  machine precision. \emph{Nat. Phys.} \textbf{2024}, \emph{20},
  1381--1382\relax
\mciteBstWouldAddEndPuncttrue
\mciteSetBstMidEndSepPunct{\mcitedefaultmidpunct}
{\mcitedefaultendpunct}{\mcitedefaultseppunct}\relax
\EndOfBibitem
\end{mcitethebibliography}

\end{document}